\documentclass[5p,times,twocolumn,11pt]{elsarticle}

\usepackage[T1]{fontenc}
\usepackage{microtype}
\usepackage{amsmath,amssymb}
\usepackage{booktabs}
\usepackage{tabularx}
\usepackage{array}
\usepackage{ragged2e}
\usepackage{enumitem}
\usepackage{placeins}
\usepackage{xcolor}
\usepackage{tikz}
\usetikzlibrary{arrows.meta,backgrounds,calc,fit,matrix,positioning}
\usepackage[colorlinks=true,citecolor=blue,linkcolor=blue,urlcolor=blue]{hyperref}
\hypersetup{
  pdftitle={A Review of Measurement System Uncertainty Handoffs Across the Manufacturing Decision Lifecycle},
  pdfauthor={Fei Jiang and Lei Yang},
  pdfkeywords={measurement uncertainty, measurement system analysis, process capability, statistical process control}
}
\biboptions{sort&compress}
\makeatletter
\g@addto@macro\ps@pprintTitle{%
  \def\@oddhead{\vbox{\hbox to\textwidth{\hfil\footnotesize\itshape Original submitted version. Accepted for publication in Measurement on November 8, 2026.\hfil}\vskip 3pt\hrule height 0.4pt}}%
  \let\@evenhead\@oddhead
  \let\@oddfoot\@empty
  \let\@evenfoot\@empty}
\makeatother

\definecolor{stageblue}{HTML}{1F5A94}
\definecolor{stageteal}{HTML}{168C8C}
\definecolor{stagegold}{HTML}{C88719}
\definecolor{stagecoral}{HTML}{B95448}
\definecolor{stageink}{HTML}{263746}
\definecolor{stagepale}{HTML}{EEF4F8}
\definecolor{abstractpink}{HTML}{D81B60}

\newcommand{\stage}[1]{\ensuremath{\mathrm{#1}}}
\newcommand{\edge}[2]{\stage{#1}$\rightarrow$\stage{#2}}
\newcommand{\chain}[3]{\stage{#1}$\rightarrow$\stage{#2}$\rightarrow$\stage{#3}}
\newcommand{\abstractcitep}[1]{\citep{#1}}

\newcolumntype{L}[1]{>{\RaggedRight\arraybackslash}p{#1}}
\newcolumntype{Y}{>{\RaggedRight\arraybackslash}X}

\begin{document}
\begin{frontmatter}

\title{A Review of Measurement System Uncertainty Handoffs Across the Manufacturing Decision Lifecycle}

\author{Fei Jiang\corref{cor1}\fnref{eqc}}
\author{Lei Yang\fnref{eqc}}
\cortext[cor1]{Corresponding author: Fei Jiang (\href{mailto:jiangfeicq@gmail.com}{jiangfeicq@gmail.com}).}
\fntext[eqc]{Fei Jiang and Lei Yang contributed equally to this work.}
\address{Independent Researcher, Seattle, WA, USA}

\begin{abstract}
Measurement systems generate uncertain information that downstream manufacturing decisions often consume as if fixed. This structured integrative review examines how measurement error, uncertainty, GR\&R estimates, calibration evidence, and drift information are handed off into capability inference, retrospective Phase I reference estimation, prospective Phase II monitoring, operational action, and requalification. Studies are compared by the upstream uncertain object, whether it is known, calibrated, or estimated, its propagation mechanism, the downstream decision, and the validation boundary. The synthesis distinguishes recorded production data from latent-process targets: measurement-system information calibrates limits, uncertainty statements, replication, guard bands, or losses, but does not reveal the realized random error in an individual record. Within the retained literature, measurement-to-capability and measurement-to-monitoring mechanisms are more extensively represented by model-based studies, whereas prospective evidence is thinner where statistical results must support source attribution, cost-sensitive action, and requalification. A decision-handoff spine separates measurement generation, statistical inference, and operational validation so that capability, stability, attribution, and intervention effectiveness are not conflated. The review is integrative rather than exhaustive. Its contribution is an auditable measurement-science framework that preserves provenance, information status, validity conditions, and downstream use while identifying claims that are model-supported, conditionally calibrated, or not yet operationally validated.
\end{abstract}


\end{frontmatter}

\section{Introduction}
\label{sec:introduction}

Measurement systems are upstream evidence-generating systems: they produce recorded values together with information about error, uncertainty, repeatability, reproducibility, calibration, and drift. Downstream manufacturing decisions consume that evidence, often after its estimated status has been obscured. Measurement system analysis (MSA) is the umbrella assessment: a gage repeatability and reproducibility (GR\&R) study estimates repeatability and reproducibility components \citep{montgomery_runger_1993a,montgomery_runger_1993b}, whereas calibration and check-standard programs characterize correction, uncertainty, and temporal credibility \abstractcitep{jcgm_100_2008,song_vorburger_2007}. These objects are related but not interchangeable under the \emph{International Vocabulary of Metrology} (VIM) and the \emph{Guide to the Expression of Uncertainty in Measurement} (GUM). Their downstream consequences depend on what information is carried forward and which decision it is intended to protect.

The manufacturing quality lifecycle supplies the decision context. In this review, \emph{Phase I} denotes retrospective screening and estimation used to establish an in-control reference state; fitted parameters and limits remain uncertain because they come from finite reference data. Capability analysis relates a declared latent-process or recorded-output target to specifications and decision risk. \emph{Phase II} denotes prospective monitoring of future recorded observations against the Phase I reference state \citep{jensen_etal_2006,psarakis_etal_2013}. Operational action and requalification then ask what intervention follows, what evidence attributes the source, and whether the affected measurement or process baseline remains fit for use.

Throughout this review, measurement error is carried into a decision rather than removed observation by observation. Operational rules act on the recorded values produced by the measurement process. Measurement-system information may alter limits, uncertainty statements, guard bands, replication plans, or loss calculations, but it does not reveal the realized random error attached to an individual reading \citep{bordignon_scagliarini_2002,pearn_liao_2005,linna_woodall_2001}. A latent process quantity may be estimated for diagnosis when the observation model and supporting information identify it; that estimate is not substituted silently for the recorded production stream.

Process-capability research has developed sampling theory for process capability indices (PCIs), including $C_p$, $C_{pk}$, and related families \citep{pearn_kotz_johnson_1992,kotz_johnson_2002,anis_2008}. Multivariate extensions and reviews emphasize that the estimand, distributional model, and inferential method constrain interpretation \citep{wu_pearn_kotz_2009,de_felipe_2017,perils_2025}, while bibliographic syntheses document the field's continued expansion \citep{yum_2011,yum_2023_part1,yum_2023_part2}.

Measurement-system assessment has separate designed-experiment, variance-component, and review foundations \citep{burdick_2003}. Recent inferential, review, and resampling work continues that line \citep{lashkari_chenouri_2025,soares_etal_2022,razumic_2025,petitjean_musset_2026}. These mature literatures establish the component tools; the unresolved synthesis question is whether their uncertain outputs are carried coherently when a later capability, monitoring, action, or requalification decision uses them.

The individual components are mature, but their uncertainty handoffs are unevenly connected. Measurement-system criteria compare estimated measurement variation with process variation or tolerance width, and different criteria need not encode the same operational risk \citep{woodall_borror_2008}. Additive measurement error can change the sampling behavior of estimated PCIs \citep{bordignon_scagliarini_2002} and weaken inference based on an observed capability-index value, particularly $C_{pk}$, unless the error model is incorporated \citep{pearn_liao_2005}. It also changes the operating characteristics of charts designed for recorded process data \citep{maravelakis_panaretos_psarakis_2004}. These pairwise results do not by themselves explain how uncertainty should be carried through capability, monitoring, action, and requalification.

A parallel literature studies uncertainty created during Phase I. Control limits are generally estimated from a reference sample rather than calculated from known process parameters. Consequently, the run-length behavior of a chart with estimated limits can differ from its known-parameter design. This distinction has been demonstrated for Shewhart charts \citep{quesenberry_1993,albers_kallenberg_2004} and synthesized across a broad range of chart families \citep{jensen_etal_2006,psarakis_etal_2013}. Estimation also enters both control-limit construction and capability analysis through common quantities such as the process standard deviation \citep{mahmoud_etal_2010}. The relevant uncertainty therefore does not belong exclusively to capability analysis or control-chart methodology; it originates upstream and may influence more than one later decision.

Some studies move beyond a single pairwise effect. Capability requirements have been incorporated into the design of acceptance-oriented control charts \citep{oprime_etal_2019}, measurement uncertainty has been connected to choices about whether to improve the measurement system or the production process \citep{villeta_etal_2012}, and check-standard histories have been used to examine whether selected components of a measurement-uncertainty budget remain credible \citep{song_vorburger_2007}. These contributions show that feedback is possible, but they also illustrate why unlike decisions must be kept separate. Acceptance relative to specifications is not identical to monitoring stability; a chart used to verify an uncertainty budget is not automatically a production-process chart; and an illustrative allocation rule does not establish a validated end-to-end control policy.

This review therefore focuses on the \emph{handoff} rather than on any one quality tool. Its central question is:

\begin{quote}
What measurement-system information is required downstream, how is its uncertainty propagated or silently treated as fixed in capability and monitoring decisions, when are latent-process diagnostics identifiable, and where does validation stop before attributed action and requalification?
\end{quote}

This review makes four related synthesis contributions. First, it treats measurement-system information as the organizing upstream object and reads each claim as a directed handoff comprising an uncertain quantity, its known/calibrated/estimated status, a propagation rule, a downstream decision, and a validation boundary. Second, it separates a rule's recorded-data input from the latent or recorded quantity it targets, so measurement information calibrates decision risk without pretending to remove realized random error from individual observations. Third, it distinguishes physical and measurement generation, statistical inference, and decision and validation evidence so that statistical performance is not relabeled as operational effectiveness. Finally, within the evidence examined here, it shows that pairwise model-based mechanisms are more extensively represented, whereas attributable, cost-sensitive action and requalification are less frequently supported by prospective evidence. Together, these distinctions define an auditable industrial information architecture in which each downstream procedure identifies the measurement-evidence version, validity domain, and uncertainty status that it consumed; they do not constitute a universal factory algorithm or standard.

These contributions are deliberately bounded at the synthesis level. The review does not claim that MSA, capability indices, parameter-estimation effects, or control-chart corrections are new, and it does not introduce a finite-GR\&R theory or a unified lifecycle likelihood. Its positioning is broader across decisions but shallower within each component than existing segment reviews: one review organizes measurement-error effects in process monitoring, another organizes GR\&R methods, and a third covers acceptance and capability-index charts. Table~\ref{tab:prior_crosswalk} summarizes the resulting novelty boundary. These reviews orient coverage but are not treated as independent replications of the primary studies they synthesize.

\begin{table*}[htbp]
\centering
\small
\caption{Novelty crosswalk against the closest review-level and normative precedents.}
\label{tab:prior_crosswalk}
\renewcommand{\arraystretch}{1.10}
\hyphenpenalty=10000\relax
\exhyphenpenalty=10000\relax
\begin{tabularx}{\textwidth}{@{}L{0.15\textwidth} L{0.26\textwidth} Y@{}}
\toprule
\textbf{Source} & \textbf{Established contribution} & \textbf{Additional inference enabled by this review} \\
\midrule
Maleki et al. \citep{maleki_2017_review} & Monitoring effects and remedies under measurement error. & Traces measurement evidence beyond monitoring into capability, action, and requalification while retaining its information status. \\
Soares et al. \citep{soares_etal_2022} & GR\&R methods and applications. & Examines how finite measurement-system estimates are consumed, treated as fixed, or propagated by downstream decisions. \\
Lizarelli et al. \abstractcitep{lizarelli_etal_2018} & Acceptance and capability-index charts. & Adds measurement provenance, estimated-input status, source-attribution boundaries, and requalification. \\
ISO/JCGM \mbox{\abstractcitep{iso_22514_4_2016,iso_22514_7_2021,iso_14253_1_2017,jcgm_100_2008,jcgm_106_2012}} & Metrology, capability, uncertainty, and conformity. & Maps normative quantities to evidence across successive decisions without treating standards as operational validation. \\
\midrule
Present review & Directed uncertainty handoffs across the decision lifecycle. & Links upstream quantity, information status, propagation, decision, and validation boundary as one auditable synthesis unit. \\
\bottomrule
\end{tabularx}
\end{table*}

The remainder of the review follows one repeated argument rather than a sequence of disconnected tool surveys: identify the uncertain upstream evidence, trace its effect on a statistical object, state the operating consequence, name the action the evidence can support, and preserve the information needed to validate or requalify that action. Section~\ref{sec:method} defines the evidence method, and Section~\ref{sec:taxonomy} formalizes this handoff spine. Sections~\ref{sec:measurement_capability}--\ref{sec:phase1_phase2} trace uncertainty forward from measurement and reference estimation to capability and monitoring. Section~\ref{sec:capability_monitoring} shows where those branches rejoin without collapsing conformity into stability. Section~\ref{sec:action_requalification} then asks when a statistical result becomes an attributable action and a defensible feedback loop. Section~\ref{sec:evidence_strength} compares evidence strength along the same spine, and Sections~\ref{sec:agenda} and \ref{sec:conclusion} state the resulting agenda and conclusions.

\section{Review scope and synthesis approach}
\label{sec:method}

\subsection{Review design and unit of analysis}

This study is a structured integrative review rather than an exhaustive systematic review. It connects established literatures around a common decision problem while preserving source assumptions and validation boundaries. The primary context is a univariate continuous manufacturing characteristic measured by a repeatable or periodically verifiable system. Multivariate methods, attribute inspection, reliability, acceptance sampling, predictive maintenance, and machine-learning diagnosis enter only when they clarify a central cross-stage mechanism or identification boundary. The unit of analysis is a \emph{directed uncertainty handoff}: an explicit connection from an upstream uncertain quantity to a downstream statistic, operating characteristic, threshold, classification, or action. Applying a control chart and then reporting a capability index is adjacent unless uncertainty from one stage is carried into the other decision.

\subsection{Corpus construction and evidence interpretation}

Sources were identified through targeted searches around measurement error and capability, measurement error and process monitoring, estimated Phase I limits, measurement uncertainty and conformity, and source attribution or requalification. DOI and publisher records were used to verify identity and publication status. Backward tracing from anchor studies and reviews, forward citation discovery, and full-text assessment were used to connect the component literatures. Search and citation tracing were last updated in August 2026, and the working route was limited to English-language records. One intellectual work was retained once, with the final peer-reviewed version controlling over duplicates or earlier versions.

Eligibility required an explicit analytical, simulation, inferential, procedural, or empirical connection from an upstream uncertain quantity to a downstream statistic, operating characteristic, threshold, classification, or action. Papers that merely applied adjacent quality tools without transferring uncertainty were treated as context rather than direct handoff evidence. Each claim was assigned to the narrowest supported handoff; a compound handoff was retained only when full text supported each connection separately. Reviews supplied coverage and discovery routes, and standards supplied terminology or normative context, but neither was counted as an independent replication. Because this was not a database-complete systematic search, the synthesis does not estimate literature prevalence or establish field-wide absence.

Evidence was interpreted by function rather than collapsed into a scalar score: analytical derivation and simulation establish model-bounded mechanisms, applied illustrations demonstrate implementation, independent empirical or prospective studies test transfer, and reviews or standards supply context. No meta-analysis was undertaken because the reported outcomes are not commensurate. Screening and interpretation used one primary review pass rather than independent duplicate review; conclusions therefore remain bounded by the retained literature and by the models and validation settings of the cited sources. Ambiguous compound handoffs were rechecked against the retained full texts, but these checks were not independent recoding, and no inter-coder reliability statistic was computed.

\section{Lifecycle taxonomy}
\label{sec:taxonomy}

\subsection{Lifecycle nodes}

The taxonomy uses six node codes: measurement qualification ($\stage{M}$), Phase I reference estimation ($\stage{I}$), capability inference ($\stage{C}$), Phase II monitoring ($\stage{II}$), operational action ($\stage{A}$), and requalification ($\stage{R}$). They describe the role of evidence in a decision, not necessarily six departments or six chronological steps.

\begin{table*}[htbp]
\centering
\caption{Normalized lifecycle nodes used in the review.}
\label{tab:nodes}
\small
\renewcommand{\arraystretch}{1.10}
\setlength{\tabcolsep}{4pt}
\hyphenpenalty=10000\relax
\exhyphenpenalty=10000\relax
\begin{tabularx}{\textwidth}{@{}>{\centering\arraybackslash}p{0.05\textwidth} L{0.22\textwidth} Y@{}}
\toprule
\textbf{Code} & \textbf{Node} & \textbf{Included evidence or decision objects} \\
\midrule
M & Measurement qualification & Error, uncertainty, bias, random variation, GR\&R, calibration, drift, and task adequacy. \\
I & Phase I estimation & Baseline screening and estimation of location, scale, distribution, and control limits. \\
C & Capability inference & Declared distributional model; estimation, interval inference, testing, classification, or acceptance relative to engineering specifications. \\
II & Phase II monitoring & False-alarm, delay, run-length, and online-decision evidence. \\
A & Operational action & Release, stop, investigate, confirm, recalibrate, improve, repair, or allocate resources. \\
R & Requalification & Checks whether the measurement system, baseline, or capability remains fit for use. \\
\bottomrule
\end{tabularx}
\end{table*}

The broad \stage{M} node is a routing category, not a claim that its quantities are interchangeable. Table~\ref{tab:nodes} supplies the lifecycle terminology used throughout the review. Table~\ref{tab:measurement_terms} then distinguishes the quantities routed through node \stage{M}. Its definitions follow the VIM/GUM distinction between an error associated with a realized quantity value and uncertainty as a non-negative parameter describing dispersion, while International Organization for Standardization (ISO) 22514-7 supplies the task-specific measurement-process capability context \abstractcitep{jcgm_200_2012,jcgm_100_2008,iso_22514_7_2021}.

\begin{table*}[htbp]
\centering
\small
\caption{Core measurement terminology and its permitted role in downstream decisions. R\&R denotes repeatability and reproducibility.}
\label{tab:measurement_terms}
\renewcommand{\arraystretch}{1.10}
\setlength{\tabcolsep}{4pt}
\hyphenpenalty=10000\relax
\exhyphenpenalty=10000\relax
\begin{tabularx}{\textwidth}{@{}L{0.17\textwidth} L{0.35\textwidth} Y@{}}
\toprule
\textbf{Quantity} & \textbf{Meaning in this review} & \textbf{Permitted propagation or claim} \\
\midrule
Measurement error & Measured value minus its reference value; may contain random and/or systematic components. & May shift location, inflate variation, or alter classification under a stated observation model; it is not synonymous with uncertainty. \\
\addlinespace[2pt]
Measurement uncertainty & Non-negative parameter characterizing the dispersion of quantity values attributed to the measurand. & Supports uncertainty budgets, conformity probabilities, guard bands, and measurement-task capability \abstractcitep{jcgm_100_2008,jcgm_106_2012}. \\
\addlinespace[2pt]
Gage R\&R & Finite-study repeatability and reproducibility variance estimates. & Carry design and estimation uncertainty; a plug-in estimate remains estimated. \\
\addlinespace[2pt]
Calibration/drift & Reference-based correction, bias, drift, and temporal-change evidence. & Supports conformity or attribution; replication cannot resolve common bias or the source of drift. \\
\addlinespace[2pt]
Measurement-system capability & Task-specific adequacy expressed by figures or ratios under the applicable standard. & Qualifies a measurement task; it is not an empirical effect estimate for process capability or monitoring \abstractcitep{iso_22514_7_2021}. \\
\bottomrule
\end{tabularx}
\end{table*}

The permitted propagation in Table~\ref{tab:measurement_terms} does not imply observation-level de-noising. A stable bias supported by calibration may be corrected, but the corrected reading retains residual uncertainty. Random measurement error is not observable for an individual production reading and therefore cannot be subtracted from that reading as if it were known. The operational record remains either the measured value or a documented calibrated value, while variance decomposition and latent-process indices remain population-level, model-dependent diagnostic summaries.

Common measurement-system ratios also answer different questions. Let $\sigma_P^2$ denote part-to-part variance, $\sigma_M^2$ measurement-system variance, $\sigma_T^2=\sigma_P^2+\sigma_M^2$, $\gamma=\sigma_P^2/\sigma_M^2$, and $\rho_P=\sigma_P^2/\sigma_T^2$. Let USL and LSL denote the upper and lower specification limits, respectively, and let $W=\mathrm{USL}-\mathrm{LSL}$ denote the specification width. A compact crosswalk of conventions discussed by Woodall and Borror is \citep{woodall_borror_2008}
\begin{equation}
\begin{gathered}
R_{P/T}=\frac{k\sigma_M}{W}
\quad
\mathrm{SNR}_{\mathrm{WB}}=\sqrt{2\gamma}
\quad
\mathrm{SNR}_{\mathrm{AIAG}}=\sqrt{\gamma},\\[2pt]
ndc^{\star}=1.41\sqrt{\gamma}
\quad
D_R=\sqrt{\frac{1+\rho_P}{1-\rho_P}}=\sqrt{2\gamma+1}.
\end{gathered}
\label{eq:measurement_metric_crosswalk}
\end{equation}
Here $R_{P/T}$ is the precision-to-tolerance ratio, SNR denotes signal-to-noise ratio, the subscript WB denotes the Woodall--Borror convention, AIAG denotes the Automotive Industry Action Group convention, $ndc$ denotes the number of distinct categories, and $D_R$ is the discrimination ratio. The AIAG convention and measurement-system guidance are documented in the fourth-edition MSA manual \abstractcitep{aiag_msa_2010}. The multiplier $k$ is the declared coverage multiplier in the precision-to-tolerance convention, and $ndc^{\star}$ is the continuous precursor before integer truncation. Thus $R_{P/T}$ is specification-dependent, whereas SNR, $ndc$, and $D_R$ compare process and measurement variation and are algebraically related only under the displayed variance decomposition. The SNR convention and multiplier $k$ must be reported explicitly; a threshold under one metric does not become the same downstream decision under another without the process-spread, tolerance, distribution, and loss assumptions needed for that translation. Woodall and Borror further recommend examining misclassification risk because category ratios alone do not encode false acceptance or false rejection relative to specifications \citep{woodall_borror_2008}.

\subsection{Core chain and handoff representation}

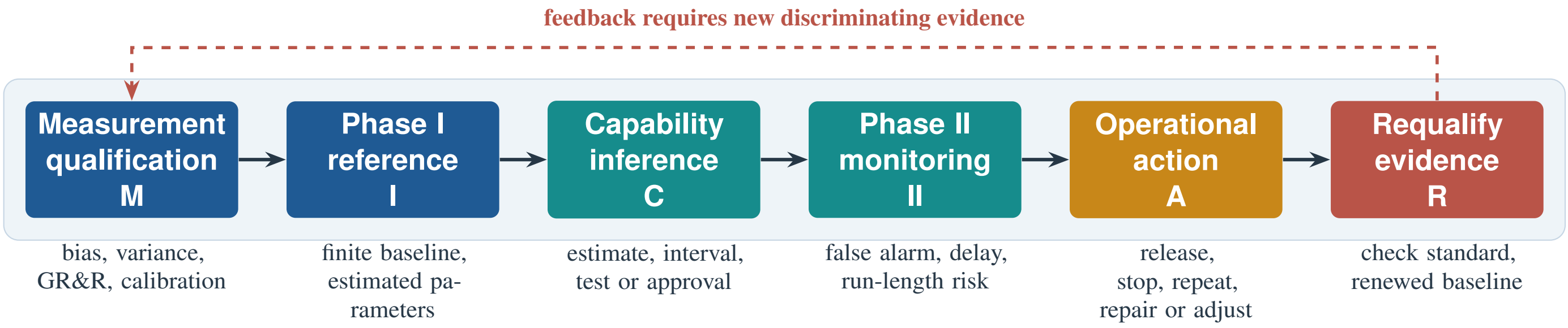
\begin{figure*}[htbp]
\centering
\resizebox{\textwidth}{!}{%
\begin{tikzpicture}[
  x=1cm,y=1cm,
  node distance=0.55cm and 0.55cm,
  stage/.style={rounded corners=3pt, minimum width=2.45cm, minimum height=1.12cm,
    text width=2.2cm, align=center, font=\sffamily\small\bfseries, text=white,
    inner sep=4pt},
  note/.style={font=\scriptsize, text=stageink, align=center, text width=2.25cm},
  flow/.style={-{Stealth[length=2.4mm]}, line width=1.15pt, draw=stageink},
  feedback/.style={-{Stealth[length=2.4mm]}, line width=1.05pt, dashed, draw=stagecoral}
]
\node[stage,fill=stageblue] (m) {Measurement\\qualification\\M};
\node[stage,fill=stageblue,right=of m] (i) {Phase I\\reference\\I};
\node[stage,fill=stageteal,right=of i] (c) {Capability\\inference\\C};
\node[stage,fill=stageteal,right=of c] (ii) {Phase II\\monitoring\\II};
\node[stage,fill=stagegold,right=of ii] (a) {Operational\\action\\A};
\node[stage,fill=stagecoral,right=of a] (r) {Requalify\\evidence\\R};

\foreach \u/\v in {m/i,i/c,c/ii,ii/a,a/r}{\draw[flow] (\u) -- (\v);}
\node[note,below=0.16cm of m] {bias, variance,\\GR\&R, calibration};
\node[note,below=0.16cm of i] {finite baseline,\\estimated parameters};
\node[note,below=0.16cm of c] {estimate, interval,\\test or approval};
\node[note,below=0.16cm of ii] {false alarm, delay,\\run-length risk};
\node[note,below=0.16cm of a] {release, stop, repeat,\\repair or adjust};
\node[note,below=0.16cm of r] {check standard,\\renewed baseline};

\coordinate (fbR) at ([yshift=0.62cm]r.north);
\coordinate (fbM) at ([yshift=0.62cm]m.north);
\draw[feedback] (r.north) -- (fbR) --
  node[above=1.5pt,font=\scriptsize\bfseries,text=stagecoral]
  {feedback requires new discriminating evidence} (fbM) -- (m.north);

\begin{scope}[on background layer]
\node[rounded corners=5pt,fill=stagepale,draw=stageblue!25,inner sep=7pt,
  fit=(m)(i)(c)(ii)(a)(r)] {};
\end{scope}
\end{tikzpicture}
}
\caption{Decision-centered lifecycle spine. The return arrow denotes feedback to measurement qualification, reference estimation, or both; it does not imply that every source validates the complete loop.}
\label{fig:lifecycle_spine}
\end{figure*}

Figure~\ref{fig:lifecycle_spine} is the spine of the review. It is read from left to right when evidence is created and used, and from right to left when an action triggers new measurement qualification or reference estimation.

Let $\mathcal D_M$ denote measurement-system information, $\widehat\psi_M$ its estimated or calibrated uncertainty description, $\mathcal D_I$ the Phase I production reference data, and $\widehat\theta_I$ the estimated process state. The review-level chain can be written as
\begin{equation}
\begin{aligned}
\mathcal D_M
&\xrightarrow{g_M}\widehat\psi_M
\longrightarrow
\mathcal D_I
\xrightarrow{g_I(\,\cdot\,;\widehat\psi_M)}\widehat\theta_I,\\
\widehat\theta_I
&\xrightarrow{g_C}\widehat C
\xrightarrow{g_{II}}S_t
\xrightarrow{g_A}a_t
\xrightarrow{g_R}r_t .
\end{aligned}
\label{eq:lifecycle_spine}
\end{equation}
Here $\widehat C$ is a capability statement whose recorded-data or latent-process estimand must be labeled, $S_t$ a statistic computed from the deployed recorded stream, $a_t$ an operational action, and $r_t$ a requalification decision. The mappings $g_M,\ldots,g_R$ represent source-specific methods spanning measurement and capability, estimated-parameter monitoring, acceptance design, and requalification \citep{bordignon_scagliarini_2002,jensen_etal_2006,oprime_etal_2019,song_vorburger_2007}. Equation~\eqref{eq:lifecycle_spine} provides the organizing spine used in this review; it does not imply that one likelihood governs the complete chain or that $g_M$ reconstructs error-free observations.

For a directed edge $j\rightarrow k$, the review records
\begin{equation}
H_{j\rightarrow k}
=\bigl(q_j,z_j,\mathcal I_j,g_{jk},d_k,V_{jk}\bigr),
\label{eq:handoff_record}
\end{equation}
where $q_j$ is the upstream uncertain quantity, $z_j$ states whether it is known, calibrated, or estimated, $\mathcal I_j$ is the available information, $g_{jk}$ is the propagation or plug-in rule, $d_k$ is the downstream decision object, and $V_{jk}$ is the validation boundary. This tuple is a review-level synthesis device, not an imported estimator: it makes unlike studies comparable without treating their numerical effects as commensurate.

The information component must be joint rather than marginal when upstream quantities are dependent. For example, if measurement drift $E_t$ and process state $X_t$ share an environmental driver or correlated dynamics, $q_j$ must retain their joint law, cross-covariance, time ordering, or an equivalent state representation; separate variances cannot identify which source produced a change in $Y_t=X_t+E_t$. Likewise, a non-Gaussian or multimodal input can be carried as a distribution, weighted sample, or source-specific simulation object rather than reduced automatically to a normal mean and variance. The handoff tuple accommodates those objects, but it does not make them identifiable: calibration artifacts, redundancy, interventions, or distinct dynamics are still required when competing sources can generate the same recorded signal.

When a differentiable downstream statistic has the form $\widehat d_k=g_{jk}(\widehat\eta_j)$, the standard multivariate delta method gives the first-order uncertainty check \abstractcitep{oehlert_1992}
\begin{equation}
\widehat{\mathrm{Var}}(\widehat d_k)
\approx
J_{g}\,\widehat\Sigma_{\eta_j}\,J_{g}^{\mathsf T},
\quad
J_g=\left.\frac{\partial g_{jk}(\eta)}{\partial\eta^{\mathsf T}}\right|_{\widehat\eta_j}.
\label{eq:generic_propagation}
\end{equation}
Equation~\eqref{eq:generic_propagation} is not imposed on studies that use exact distributions, bootstrap procedures, simulation, or non-differentiable rules. It supplies a diagnostic question: did the downstream analysis carry the uncertainty of its upstream estimate, or did it replace that estimate by a fixed plug-in value?

That question separates three treatments that are often described with the same word ``uncertainty.'' A known or externally calibrated input supports performance conditional on a supplied value. A plug-in analysis inserts a realized estimate but then treats it as fixed. A propagated analysis also averages over, or otherwise represents, the estimation distribution of that input. Table~\ref{tab:uncertainty_treatments} makes the distinction operational.

\begin{table*}[htbp]
\centering
\small
\caption{Three treatments of an upstream quantity in a downstream analysis.}
\label{tab:uncertainty_treatments}
\renewcommand{\arraystretch}{1.10}
\setlength{\tabcolsep}{4pt}
\hyphenpenalty=10000\relax
\exhyphenpenalty=10000\relax
\begin{tabularx}{\textwidth}{@{}L{0.17\textwidth} L{0.20\textwidth} Y@{}}
\toprule
\textbf{Treatment} & \textbf{Downstream calculation} & \textbf{Interpretation and required reporting} \\
\midrule
Known or calibrated & Condition on supplied $\eta$. & Report how $\eta$ was established and which calibration or model uncertainty remains outside the calculation. \\
Plug-in estimate & Condition on realized $\widehat\eta$. & Report evidence for $\widehat\eta$; performance remains conditional. \\
Propagated estimate & Integrate/simulate over $\widehat\eta$. & Report the joint information structure and propagation method, and whether the guarantee is conditional, marginal, or percentile-based. \\
\bottomrule
\end{tabularx}
\end{table*}

For a downstream random quantity $D$ whose distribution depends on an upstream estimate $\widehat\eta$, the standard law of total variance provides a method-neutral decomposition,
\begin{equation}
\mathrm{Var}(D)
=
\mathrm{E}_{\widehat\eta}\!\left\{\mathrm{Var}(D\mid\widehat\eta)\right\}
+
\mathrm{Var}_{\widehat\eta}\!\left\{\mathrm{E}(D\mid\widehat\eta)\right\}.
\label{eq:total_variance_handoff}
\end{equation}
The first term describes average downstream variation conditional on the realized input; the second describes variation across possible upstream estimates. A plug-in calculation retains the first perspective but omits the second unless estimation uncertainty is restored analytically, by resampling, or by simulation. Equation~\eqref{eq:total_variance_handoff} is a standard probability identity used here as a review diagnostic, not a new estimator and not a requirement that every source adopt a hierarchical model. Estimated-parameter chart studies illustrate why the distinction can change deployed performance \citep{jensen_etal_2006,psarakis_etal_2013}.

The final transition is decision-theoretic rather than purely statistical. Following the expected-loss logic used in uncertainty-aware conformity assessment \citep{forbes_2006,kosztyan_2017}, an operational action is fully specified only when the available information $\mathcal I_t$, state $\Omega_t$, action set $\mathcal A$, and loss $L$ determine
\begin{equation}
a_t^{\star}
=\arg\min_{a\in\mathcal A}
\mathrm E\!\left\{L(a,\Omega_t)\mid\mathcal I_t\right\}.
\label{eq:decision_endpoint}
\end{equation}
Most reviewed papers estimate a statistical operating characteristic without evaluating Equation~\eqref{eq:decision_endpoint}. This distinction separates evidence of altered bias, coverage, or run length from evidence that a particular intervention improves the manufacturing outcome.

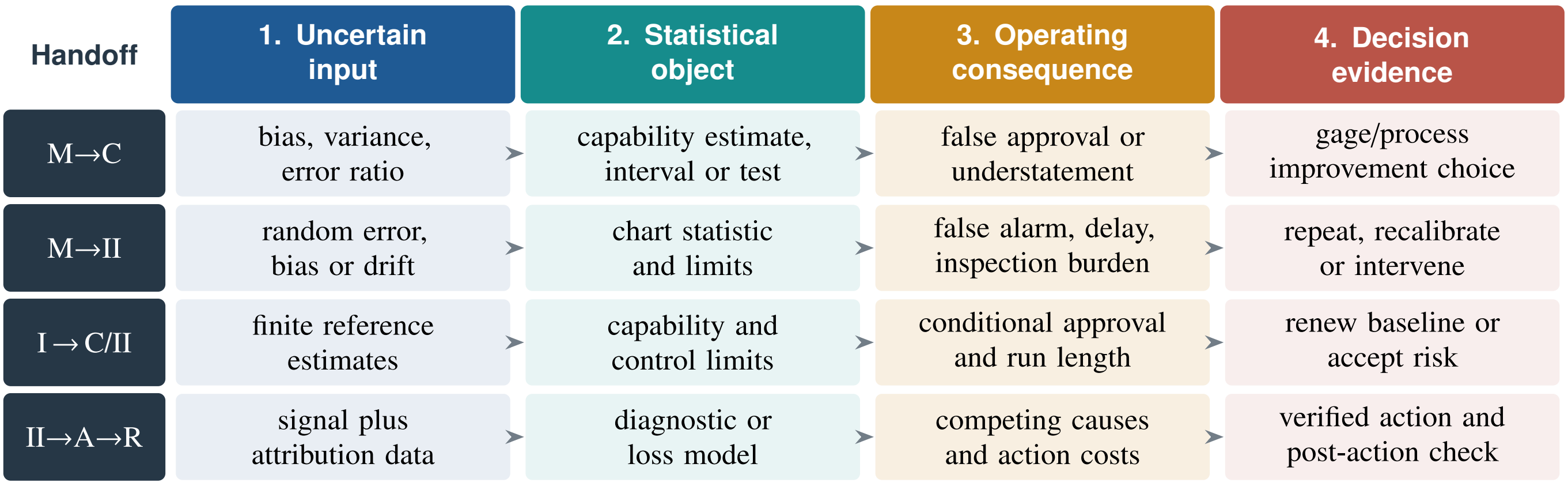
\begin{figure*}[htbp]
\centering
\resizebox{\textwidth}{!}{%
\begin{tikzpicture}[
  cell/.style={rounded corners=2pt, minimum height=0.84cm, text width=3.05cm,
    align=center, anchor=center, font=\scriptsize, inner sep=2.5pt, draw=white},
  rowlab/.style={rounded corners=2pt, minimum height=0.84cm, text width=1.3cm,
    align=center, anchor=center, font=\sffamily\scriptsize\bfseries, text=white, fill=stageink},
  head/.style={rounded corners=2pt, minimum height=0.94cm, text width=3.05cm,
    align=center, anchor=center,font=\sffamily\scriptsize\bfseries, text=white},
  arrow/.style={-{Stealth[length=1.8mm]},draw=stageink!65,line width=0.7pt}
]
\matrix (map) [matrix of nodes,ampersand replacement=\&,row sep=1.5pt,column sep=1.5pt,nodes in empty cells] {
|[rowlab,fill=white,text=stageink]|{Handoff} \& |[head,fill=stageblue]|{1. Uncertain\\input} \& |[head,fill=stageteal]|{2. Statistical\\object} \& |[head,fill=stagegold]|{3. Operating\\consequence} \& |[head,fill=stagecoral]|{4. Decision\\evidence} \\
|[rowlab]|{\edge{M}{C}} \& |[cell,fill=stageblue!10]|{bias, variance,\\error ratio} \& |[cell,fill=stageteal!10]|{capability estimate,\\interval or test} \& |[cell,fill=stagegold!13]|{false approval or\\understatement} \& |[cell,fill=stagecoral!10]|{gage/process\\improvement choice} \\
|[rowlab]|{\edge{M}{II}} \& |[cell,fill=stageblue!10]|{random error,\\bias or drift} \& |[cell,fill=stageteal!10]|{chart statistic\\and limits} \& |[cell,fill=stagegold!13]|{false alarm, delay,\\inspection burden} \& |[cell,fill=stagecoral!10]|{repeat, recalibrate\\or intervene} \\
|[rowlab]|{\stage{I}\,$\rightarrow$\,\stage{C}/\stage{II}} \& |[cell,fill=stageblue!10]|{finite reference\\estimates} \& |[cell,fill=stageteal!10]|{capability and\\control limits} \& |[cell,fill=stagegold!13]|{conditional approval\\and run length} \& |[cell,fill=stagecoral!10]|{renew baseline or\\accept risk} \\
|[rowlab]|{\chain{II}{A}{R}} \& |[cell,fill=stageblue!10]|{signal plus\\attribution data} \& |[cell,fill=stageteal!10]|{diagnostic or\\loss model} \& |[cell,fill=stagegold!13]|{competing causes\\and action costs} \& |[cell,fill=stagecoral!10]|{verified action and\\post-action check} \\
};
\foreach \rr in {2,3,4,5}{
  \draw[arrow] (map-\rr-2.east) -- (map-\rr-3.west);
  \draw[arrow] (map-\rr-3.east) -- (map-\rr-4.west);
  \draw[arrow] (map-\rr-4.east) -- (map-\rr-5.west);
}
\end{tikzpicture}
}
\caption{Argument map for the review. The same uncertainty can feed parallel statistical objects, but a statistical effect becomes a lifecycle claim only when its downstream decision and validation boundary are stated. A source supports a handoff only to the furthest decision stage that it evaluates.}
\label{fig:argument_map}
\end{figure*}

Equations~\eqref{eq:lifecycle_spine}--\eqref{eq:decision_endpoint} define the review architecture, propagation diagnostics, and decision endpoint. The remaining sections follow one repeated chain: uncertain evidence changes a statistical object; that change matters through an operating consequence; and a lifecycle claim is complete only when its supported action and validation or requalification boundary are stated. Figure~\ref{fig:argument_map} keeps that sequence visible while Sections~\ref{sec:measurement_capability}--\ref{sec:action_requalification} examine its branches; none of the review representations is treated as a newly validated estimator.

The distinction between \stage{M} and \stage{I} separates uncertainty attributable to obtaining a measurement from uncertainty created when estimating the process reference state. The observed data can contain both. Coding them separately does not assume that they are identifiable from production measurements alone. A source must state the error structure, use calibration or check-standard information, exploit replication, or otherwise provide an information structure that supports the distinction.

The distinction between \stage{C} and \stage{II} separates conformity or capability relative to specifications from temporal monitoring relative to an in-control reference state. A highly capable process can still exhibit a special-cause signal, and a statistically stable process can be incapable relative to specifications. Acceptance-oriented chart designs may connect these nodes, but that connection is recorded as a decision chain rather than used to erase the conceptual distinction \citep{oprime_etal_2019}.

Finally, \stage{A} and \stage{R} make feedback explicit. A statistical effect is not yet an action unless the source defines what is done with the result. Requalification is more specific than investigation: it asks whether a measurement system, uncertainty statement, reference baseline, or capability conclusion must be checked or renewed. The dynamic-limit check-standard method of Song and Vorburger \citep{song_vorburger_2007}, for example, informs a bounded requalification question; it does not establish the absence of measurement bias.

To keep this chain concrete, Table~\ref{tab:running_case} follows one illustrative case: a continuous dimensional characteristic is measured by a repeatable gage, summarized in a finite Phase I reference study, classified relative to specifications, monitored in Phase II, investigated after a signal, and requalified using new discriminating evidence. Later references resume this same hypothetical thread; it is not a separate dataset or a claim that the cited studies examined one common factory. The example is illustrative and is not presented as empirical validation of the lifecycle framework.

\subsection{Edge representation}

An edge is coded when evidence from one node is used to alter or evaluate a decision at another node. The main direct edges in the present review are \edge{M}{I}, \edge{M}{C}, \edge{M}{II}, \edge{I}{C}, \edge{I}{II}, \edge{II}{A}, \edge{M}{A}, and \edge{A}{R}. The expanded synthesis also retains \edge{M}{R} when a measurement-uncertainty statement is directly rechecked.

Three distinctions govern interpretation:

\begin{enumerate}[leftmargin=*]
    \item \textbf{Direct effect versus sequential chain.} If measurement error directly changes a capability estimator, the evidence is \edge{M}{C}. If the capability result then determines whether measurement or process improvement is selected, the full logic is \chain{M}{C}{A}. The second edge requires an explicit action rule.
    \item \textbf{Parallel destinations versus shared decision logic.} A scale estimator may enter both a control limit and a capability index, supporting linked but distinct destinations \citep{mahmoud_etal_2010}. By contrast, a capability-constrained acceptance chart uses the capability decision inside the monitoring design and is coded \chain{I}{C}{II}.
    \item \textbf{Monitoring the process versus monitoring the measurement system.} A chart of production measurements belongs to \stage{II} when its operating characteristics concern the production state. A chart of a check standard belongs to \stage{R} when it evaluates the continued credibility of measurement uncertainty \citep{song_vorburger_2007}.
\end{enumerate}

\begin{table*}[t]
\centering
\small
\caption{Industrial interpretation of the uncertainty-handoff framework for the illustrative dimensional case.}
\label{tab:running_case}
\renewcommand{\arraystretch}{1.10}
\setlength{\tabcolsep}{4pt}
\hyphenpenalty=10000\relax
\exhyphenpenalty=10000\relax
\begin{tabularx}{\textwidth}{@{}>{\centering\arraybackslash}p{0.045\textwidth} L{0.41\textwidth} Y@{}}
\toprule
\textbf{Code} & \textbf{Retained handoff record} & \textbf{Immediate use and inference to avoid} \\
\midrule
M & Bias, GR\&R, calibration, drift estimates; information size; interval/covariance; design; conditions; version. & Qualify the measurement task and calibrate later inference; avoid treating an estimate as known or realized error in one observation. \\
I & Recorded subgroups; $n_I$, $\widehat\mu_I$, and $\widehat\sigma_I$; estimator; screening assumptions; baseline version. & Establish the reference state; avoid treating fitted limits as known-parameter limits. \\
C & Specifications; measurement and baseline versions; estimand; spread; uncertainty treatment; criterion. & Infer capability for a labeled recorded or latent target; avoid switching targets or observation-level de-noising. \\
II & Future recorded stream; baseline and measurement versions; limits; conditional or marginal criterion. & Monitor future records; avoid interpreting a signal as identification of its physical cause. \\
A & Signal; attribution evidence; action rule; relevant loss and costs. & Investigate, contain, recalibrate, adjust, or collect evidence; avoid automatic process adjustment after an alarm. \\
R & Check, calibration, redundancy, and post-action evidence; renewed record; affected consumers. & Restore or update validity; avoid treating one successful check as universal absence of bias or drift. \\
\bottomrule
\end{tabularx}
\end{table*}

This representation prevents adjacency from being mistaken for propagation. It also keeps the unit of synthesis close to the claim a practitioner must act on. One article may contribute several claims, but each claim retains its own upstream quantity, assumptions, decision object, and validation boundary.

\subsection{Cross-classification dimensions}

Each edge is compared by its upstream uncertain object and status, sample information, information structure, propagation mode, downstream decision, evidence function, and transfer boundary. The uncertainty-status dimension is especially important. A variance component estimated from a finite GR\&R study, a calibration uncertainty reported by a laboratory, and a fixed measurement-error ratio used in a simulation are not interchangeable inputs. Similarly, a Phase I estimate based on a finite reference sample should not be discussed as if the chart parameters were known. Reviews of parameter-estimation effects provide broad evidence for this distinction \citep{jensen_etal_2006,psarakis_etal_2013}; the present taxonomy extends the same discipline to capability, action, and requalification handoffs.

The taxonomy is descriptive rather than a new statistical model. It does not impose a common likelihood, loss function, or estimator on the underlying studies. Its role is to make comparisons auditable: a later synthesis statement must identify the supported edge, the status of the upstream quantity, the downstream decision, and the evidence boundary. This structure allows mature pairwise literatures to be acknowledged while exposing where evidence for an end-to-end lifecycle policy remains incomplete.

\subsection{Industrial implementation of uncertainty handoffs}
\label{sec:industrial_implementation}

The framework can be implemented as a measurement-information handoff contract or versioned uncertainty record, not as a universal industrial algorithm. Its identity and governance fields include a handoff identifier, version, effective time, and responsible owner. Its measurement fields identify the characteristic, method, instrument or system, operating range, and environment. Its evidence fields retain whether the upstream quantity is known, calibrated, or estimated; the source experiment or dataset; design and information size; estimator; interval, covariance, distribution, or simulation representation; assumptions; validity domain; expiry rule; and downstream consumers.

Scheduled calibration, maintenance or replacement, environmental or product-family change, detected drift, process-regime change, and invalidation of the Phase I baseline are candidate requalification triggers. Each downstream procedure should reference the exact upstream record version that it consumed. Otherwise an estimated quantity can silently become a timeless known constant after crossing a software, departmental, or analytical boundary.

Some elements are directly implementable now: preserve provenance and information status, label recorded and latent estimands, retain the finite-estimation status of Phase I inputs, distinguish alarm from attribution, and version records and triggers. Model-dependent elements include capability correction, chart calibration, replication design, guard bands, and conditional or marginal guarantees. The review does not supply a universal optimal intervention, alarm-source probability, recalibration interval, or validated factory policy. It supplies an auditable information architecture and decision discipline within which source-specific procedures can be used and tested.

For a compact numerical trace, suppose the hypothetical characteristic has $\mathrm{LSL}=9.500$ and $\mathrm{USL}=10.500$. A finite measurement study records $\widehat\sigma_E=0.040$ and a sensitivity range of $0.030$--$0.050$; this range is illustrative and is not a confidence interval. A Phase I sample of $n_I=50$ gives $\widehat\mu_I=10.100$ and recorded spread $\widehat\sigma_Y=0.120$. The recorded-data calculation gives $\widehat C_{pk,Y}=1.11$. Under the independent additive, zero-bias model used later, a plug-in latent diagnostic gives $\widehat C_{pk,X}=1.18$, with sensitivity values about $1.15$--$1.22$ across the stated gage range. These are different estimands, not corrected individual measurements. In deployment, each monitoring rule should identify the approved measurement-record and Phase I-baseline versions it uses. If the rule signals, the organization first contains or investigates and obtains discriminating evidence; it does not automatically adjust the process. Evidence of gage drift would prompt service or recalibration, a new measurement-record version, and review of whether the capability statement, Phase I baseline, chart limits, release decision, or action rule has become stale. Requalification therefore renews affected evidence and consumers rather than merely ``fixing the instrument.''

\section{Measurement uncertainty and capability inference}
\label{sec:measurement_capability}

\noindent\textit{Handoff question.} For the illustrative characteristic, what measurement quantity is uncertain, how does it change the capability estimate or approval, and when can that change justify improving the gage rather than the production process? This first forward branch follows \edge{M}{C} and, only when an action rule is supplied, \chain{M}{C}{A}.

\subsection{Capability estimand and index conventions}

A capability symbol is not self-interpreting: its spread basis, centering convention, specification geometry, distributional model, and data target must travel with it. For a bilateral specification under a normal location--scale reference model, let $s>0$ denote a declared process-spread quantity or estimator and let $\mu$ denote the corresponding location target. A compact convention-neutral crosswalk is \citep{pearn_kotz_johnson_1992,kotz_johnson_2002,anis_2008,wu_pearn_kotz_2009,iso_22514_4_2016}
\begin{equation}
\begin{aligned}
C_p(s)&=\frac{\mathrm{USL}-\mathrm{LSL}}{6s},\\
C_{pk}(\mu,s)&=\min\left\{
\frac{\mathrm{USL}-\mu}{3s},
\frac{\mu-\mathrm{LSL}}{3s}
\right\}.
\end{aligned}
\label{eq:capability_convention_crosswalk}
\end{equation}
Equation~\eqref{eq:capability_convention_crosswalk} uses $C_p$ and $C_{pk}$ as generic algebraic templates rather than as labels that determine the spread estimator. When $s$ represents a within-condition or reference-state spread, the resulting quantities are commonly labeled $C_p$ and $C_{pk}$; when $s$ represents an overall spread, common performance-measure notation is $P_p$ and $P_{pk}$. The labels and estimators vary across standards and software, so the reported terminology must change with the declared definition of $s$, the subgroup or time structure, and the stability conditions used to interpret it. The two terms inside the minimum are the upper and lower one-sided components; with a unilateral specification, only the applicable component is defined rather than an artificial bilateral minimum.

The crosswalk is not a universal nonnormal formula. Nonnormal capability may be based on fitted distributions, transformations, or quantiles, and those choices need not yield interchangeable estimands. Likewise, pooled within-subgroup standard deviations, range- or moving-range-based estimators, overall sample standard deviations, and robust or model-specific scales encode different information and sampling uncertainty \citep{mahmoud_etal_2010,de_felipe_2017,perils_2025}. Every downstream handoff must therefore report the distributional method, scale estimator, bilateral or unilateral specification, and whether the target is the recorded stream or a model-identified latent process. The operational rule still acts on recorded measurements; choosing a latent-process target does not remove the realized error from an individual record.

\subsection{From a measurement study to an uncertainty object}

The later capability equations require a bias, error variance, error ratio, or uncertainty statement, but a measurement-system study does not deliver those quantities without estimation. For a balanced crossed GR\&R design with $o$ operators, $p$ parts, and $n$ repeated measurements per operator--part cell, the random-effects representation of Montgomery and Runger can be written in recorded-value notation as \citep{montgomery_runger_1993a,montgomery_runger_1993b}
\begin{equation}
Y_{ijk}
=\mu+O_i+P_j+(OP)_{ij}+\varepsilon_{ijk},
\label{eq:grr_random_effects}
\end{equation}
where $i=1,\ldots,o$, $j=1,\ldots,p$, and $k=1,\ldots,n$. In the source model, $O_i$, $P_j$, $(OP)_{ij}$, and $\varepsilon_{ijk}$ are mutually independent zero-mean normal random effects with variances $\sigma_O^2$, $\sigma_P^2$, $\sigma_{OP}^2$, and $\sigma_R^2$, respectively. Here $\sigma_R^2$ is the within-cell repeatability variance; the part component $\sigma_P^2$ describes variation among sampled parts and is not a gage component. The measurement components are
\begin{equation}
\begin{aligned}
\sigma_{\mathrm{repeat}}^2&=\sigma_R^2,\\
\sigma_{\mathrm{repro}}^2&=\sigma_O^2+\sigma_{OP}^2,\\
\sigma_{\mathrm{gage}}^2&=\sigma_{\mathrm{repeat}}^2+\sigma_{\mathrm{repro}}^2.
\end{aligned}
\label{eq:grr_variance_decomposition}
\end{equation}
For the full balanced model, the source's analysis-of-variance (ANOVA) moment estimators can be expressed as \citep{montgomery_runger_1993b}
\begin{equation}
\begin{aligned}
\widehat\sigma_R^2&=MS_R,\\
\widehat\sigma_{OP}^2&=\frac{MS_{OP}-MS_R}{n},\\
\widehat\sigma_O^2&=\frac{MS_O-MS_{OP}}{pn}.
\end{aligned}
\label{eq:grr_anova_estimators}
\end{equation}
where $MS_R$, $MS_{OP}$, and $MS_O$ denote the repeatability, operator--part interaction, and operator mean squares; here $MS$ means mean square, not measurement system. Consequently, $\widehat\sigma_{\mathrm{repro}}^2=\widehat\sigma_O^2+\widehat\sigma_{OP}^2$ and $\widehat\sigma_{\mathrm{gage}}^2=\widehat\sigma_R^2+\widehat\sigma_{\mathrm{repro}}^2$ are estimates rather than known error parameters. Under the stated normal full-model assumptions, $\nu_R=op(n-1)$ is the repeatability error degrees of freedom and an exact $100(1-\alpha)\%$ interval is \citep{montgomery_runger_1993b}
\begin{equation}
\left[
\frac{\nu_R\widehat\sigma_R^2}{\chi^2_{\nu_R,\,1-\alpha/2}},
\frac{\nu_R\widehat\sigma_R^2}{\chi^2_{\nu_R,\,\alpha/2}}
\right],
\label{eq:grr_repeatability_ci}
\end{equation}
where $\chi^2_{\nu,q}$ is the $q$ quantile of a chi-square distribution with $\nu$ degrees of freedom. Corresponding intervals for reproducibility and total gage variance generally require source-specific approximations or other inferential methods; they do not inherit Equation~\eqref{eq:grr_repeatability_ci} unchanged \citep{montgomery_runger_1993b,burdick_2003,lashkari_chenouri_2025}.

Equations~\eqref{eq:grr_random_effects}--\eqref{eq:grr_repeatability_ci} supply a technical entrance to the lifecycle, not a new finite-GR\&R method. Interaction, imbalance, nested factors, destructive measurement, within-part variation, dependence, and the choice among ANOVA, likelihood, generalized, bootstrap, or Bayesian inference can change the valid model and uncertainty statement \citep{burdick_2003,majeske_2012,soares_etal_2022,lashkari_chenouri_2025}. Moment estimators can also be negative at a variance-component boundary; replacing them by zero, deleting a term, or using a constrained estimator changes the inferential procedure and must be reported. The handoff object $\widehat\psi_M$ should therefore carry the design, point estimates, interval or covariance information, and estimation method. A task-specific $\widehat\sigma_{\mathrm{gage}}^2$ may inform the $\sigma_E^2$ used below only when the observation model, scale, temporal regime, and included components agree; the symbols are not automatically interchangeable.

\subsection{Recorded-data decisions and latent-process targets}

Capability indices compare the location and spread of a process with specification limits, but the required process quantities are inferred from measured data. When measurement error contributes to the observed spread, an index calculated from that spread does not describe the latent production process alone. This is the central \edge{M}{C} handoff: an upstream representation of measurement performance changes a downstream estimate, interval, test, or capability classification.

The mechanism can be stated compactly using the additive independent-error models employed in capability and chart studies \citep{mittag_1997,bordignon_scagliarini_2002,linna_woodall_2001}. Let the latent product characteristic be $X\sim \mathcal{N}(\mu,\sigma_X^2)$ and let the measurement error be $E\sim \mathcal{N}(b,\sigma_E^2)$, independent of $X$; allowing $b\neq0$ makes the bias term explicit. The recorded value is
\begin{equation}
Y=X+E,
\quad
\mathrm{E}(Y)=\mu+b,
\quad
\mathrm{Var}(Y)=\sigma_X^2+\sigma_E^2.
\label{eq:additive_measurement_model}
\end{equation}
Using the specification width $W$ defined in Section~\ref{sec:taxonomy}, let $m=(\mathrm{USL}+\mathrm{LSL})/2$ and $\tau=\sigma_E/\sigma_X$. The conventional latent-process indices are \citep{pearn_kotz_johnson_1992,kotz_johnson_2002}
\begin{equation}
C_{p,X}=\frac{W}{6\sigma_X},
\quad
C_{pk,X}=\frac{W/2-|\mu-m|}{3\sigma_X}.
\label{eq:latent_capability}
\end{equation}
If the same definitions are applied to the recorded values, the additive-error relation gives \citep{mittag_1997,bordignon_scagliarini_2002}
\begin{equation}
C_{p,Y}=\frac{C_{p,X}}{\sqrt{1+\tau^2}},
\quad
C_{pk,Y}=\frac{W/2-|\mu+b-m|}
{3\sigma_X\sqrt{1+\tau^2}}.
\label{eq:observed_capability}
\end{equation}
For zero-mean error ($b=0$), both observed indices are attenuated by $1/\sqrt{1+\tau^2}$ under this model \citep{bordignon_scagliarini_2002}. Equation~\eqref{eq:observed_capability} also shows why bias and random measurement variation must not be merged: bias changes the numerator of $C_{pk}$, whereas random error inflates the denominator of both indices.

Equations~\eqref{eq:additive_measurement_model}--\eqref{eq:observed_capability} distinguish an estimand from the data used to make a decision. A capability decision may target the latent process quantity $C_{pk,X}$ or the recorded-output quantity $C_{pk,Y}$, but the implemented rule still receives recorded observations $Y_1,\ldots,Y_n$ and measurement-system information $\mathcal D_M$. Under an identified model it may infer $C_{pk,X}$ and calibrate false-approval risk; it does not recover the realized $X_i$ by subtracting an unobserved random error $E_i$ from each $Y_i$. A documented calibration correction for stable bias changes the recorded value and must carry residual calibration uncertainty forward. This separation is essential: population-level variance decomposition can support a latent-process statement, while release, monitoring, and traceability continue to operate on the documented measurement record.

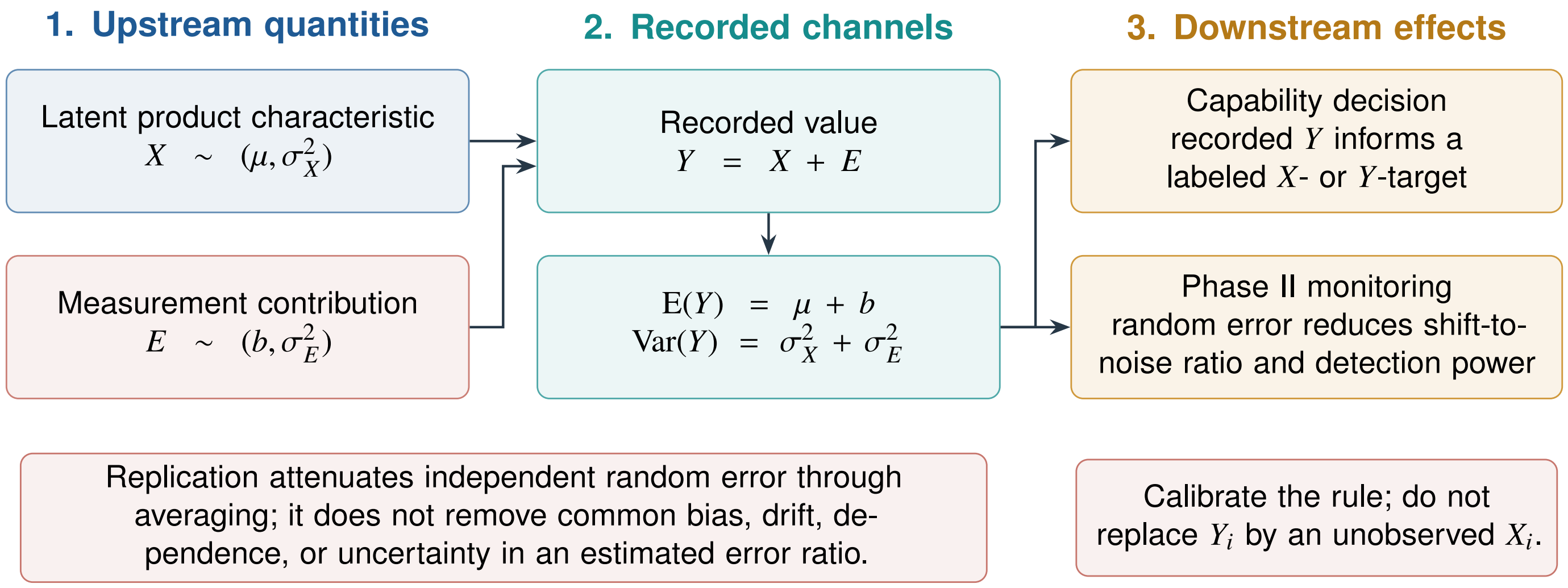
\begin{figure*}[htbp]
\centering
\resizebox{\textwidth}{!}{%
\begin{tikzpicture}[
  font=\sffamily\scriptsize,
  source/.style={rounded corners=3pt,draw=stageblue!70,fill=stageblue!8,
    minimum width=4.05cm,minimum height=1.25cm,text width=3.75cm,align=center,inner sep=3pt},
  channel/.style={rounded corners=3pt,draw=stageteal!75,fill=stageteal!8,
    minimum width=4.05cm,minimum height=1.25cm,text width=3.75cm,align=center,inner sep=3pt},
  outcome/.style={rounded corners=3pt,draw=stagegold!85,fill=stagegold!10,
    minimum width=4.30cm,minimum height=1.25cm,text width=4.0cm,align=center,inner sep=3pt},
  warn/.style={rounded corners=3pt,draw=stagecoral!80,fill=stagecoral!8,
    minimum height=1.02cm,text width=4.0cm,align=center,font=\sffamily\scriptsize,inner sep=3pt},
  head/.style={minimum width=4.05cm,minimum height=0.44cm,text width=3.95cm,
    align=center,font=\sffamily\footnotesize\bfseries},
  arr/.style={-{Stealth[length=1.8mm]},draw=stageink,line width=0.62pt}
]
\node[source] (x) at (0,2.45) {Latent product characteristic\\$X\sim(\mu,\sigma_X^2)$};
\node[source,draw=stagecoral!75,fill=stagecoral!8] (e) at (0,0.82)
  {Measurement contribution\\$E\sim(b,\sigma_E^2)$};
\node[channel] (y) at (4.65,2.45)
  {Recorded value\\$Y=X+E$};
\node[channel] (mom) at (4.65,0.82)
  {$\mathrm E(Y)=\mu+b$\\$\mathrm{Var}(Y)=\sigma_X^2+\sigma_E^2$};

\node[outcome] (cap) at (9.45,2.45)
  {Capability decision\\recorded $Y$ informs a labeled $X$- or $Y$-target};
\node[outcome] (mon) at (9.45,0.82)
  {Phase II monitoring\\random error reduces shift-to-noise ratio and detection power};
\node[warn] (dec) at (9.45,-0.85)
  {Calibrate the rule; do not replace $Y_i$ by an unobserved $X_i$.};

\node[head,text=stageblue] at (0,3.45) {1. Upstream quantities};
\node[head,text=stageteal] at (4.65,3.45) {2. Recorded channels};
\node[head,text=stagegold!90!black] at (9.45,3.45) {3. Downstream effects};

\coordinate (eybend) at ($(e.east)!0.5!(e.east -| y.west)$);
\coordinate (mcbend) at ($(mom.east)!0.5!(mom.east -| cap.west)$);
\draw[arr] (x.east) -- (y.west);
\draw[arr] (e.east) -- (eybend) |- ([yshift=-0.22cm]y.west);
\draw[arr] (y.south) -- (mom.north);
\draw[arr] (mom.east) -- (mcbend) |- (cap.west);
\draw[arr] (mom.east) -- (mon.west);

\node[warn,text width=8.25cm] at (2.33,-0.85)
  {Replication attenuates independent random error through averaging; it does not remove common bias, drift, dependence, or uncertainty in an estimated error ratio.};
\end{tikzpicture}
}
\caption{Measurement-error propagation anatomy. The diagram distinguishes latent process variation, measurement bias and random error, their recorded-data moments, and capability and monitoring consequences under source-specific assumptions. Measurement-system information calibrates an inference or decision rule; it does not remove realized random error from individual observations.}
\label{fig:measurement_propagation}
\end{figure*}

Figure~\ref{fig:measurement_propagation} turns those algebraic distinctions into a source-bounded mechanism map. It follows the additive-error capability results and the control-chart power mechanism in the cited literature, while keeping a supplied error parameter separate from an error component estimated in a finite measurement-system study \citep{mittag_1997,bordignon_scagliarini_2002,linna_woodall_2001,celano_2022,dalalah_hani_2016}.

The effect is target-sensitive for $C_{pm}$ rather than being a universal copy of the $C_p$ attenuation. Let $T_0$ denote the target. Under the same independent normal-error model, the latent and recorded-data forms are \citep{bordignon_2006}
\begin{equation}
\begin{aligned}
C_{pm,X}&=\frac{W}{6\sqrt{\sigma_X^2+(\mu-T_0)^2}},\\
C_{pm,Y}&=\frac{W}{6\sqrt{\sigma_X^2+\sigma_E^2+(\mu+b-T_0)^2}}.
\end{aligned}
\label{eq:cpm_observed}
\end{equation}
For $b=0$, their ratio is $C_{pm,Y}/C_{pm,X}=\sqrt{(1+A^2)/(1+A^2+\tau^2)}$, where $A=|\mu-T_0|/\sigma_X$. Bordignon and Scagliarini further show that finite-sample bias and mean-squared error depend on $n$, $A$, and $\tau$, and illustrate how a crossed GR\&R study can estimate the contamination ratio \citep{bordignon_2006}. The source does not propagate uncertainty in that estimated ratio through a final operational decision.

The clearest analytical studies begin with additive independent measurement error under normality. In this setting, the recorded variance contains both process and measurement components. Bordignon and Scagliarini derived sampling properties for estimators of $C_p$ and $C_{pk}$ when observations are contaminated by measurement error \citep{bordignon_scagliarini_2002}. Their analysis separates sampling error from measurement error and shows why a smaller estimator variance is not necessarily evidence of better inference: contamination can move the estimator away from the desired process-only index, so mean squared error can increase even when the sampling distribution appears more concentrated. The practical implication is bounded but important. A capability estimate calculated from recorded data cannot be interpreted as a conservative or precise estimate of latent process capability without specifying the measurement-error model and target quantity; model-based inference for that target does not create error-free individual observations.

Their $C_p$ result also exposes the interaction between finite-sample bias and measurement contamination. Let $S_Y$ be the usual sample standard deviation of $n$ recorded observations, let $f=n-1>1$, and let $\sigma_Y=\sigma_X\sqrt{1+\tau^2}$. Because $fS_Y^2/\sigma_Y^2\sim\chi_f^2$, the reciprocal moment needed by $\widehat C_{p,Y}=W/(6S_Y)$ gives the Bordignon--Scagliarini factor \citep{bordignon_scagliarini_2002}
\begin{equation}
b_f=\sqrt{\frac{2}{f}}\,
\frac{\Gamma(f/2)}{\Gamma\{(f-1)/2\}},
\quad f>1.
\label{eq:c4_factor}
\end{equation}
$b_f$ is a dimensionless reciprocal-scale finite-sample factor arising from $\mathrm E(1/S_Y)$; it separates bias due to estimating an inverse standard deviation from attenuation due to measurement contamination. It is not a measurement-error ratio: $b_f<1$ and approaches one as the degrees of freedom increase, whereas contamination enters separately through $(1+\tau^2)^{-1/2}$.
Indeed, if $Q=fS_Y^2/\sigma_Y^2$, then $Q\sim\chi_f^2$ and
\begin{equation}
\mathrm{E}\!\left(\frac{1}{S_Y}\right)
=\frac{\sqrt f}{\sigma_Y}\,\mathrm{E}(Q^{-1/2})
=\frac{1}{b_f\sigma_Y}.
\label{eq:reciprocal_sample_sd}
\end{equation}
Consequently, under the independent normal model,
\begin{equation}
\begin{aligned}
\mathrm{E}(\widehat C_{p,Y})
&=\frac{C_{p,X}}{b_f\sqrt{1+\tau^2}},\\
\mathrm{Bias}_{C_{p,X}}(\widehat C_{p,Y})
&=C_{p,X}\left\{\frac{1}{b_f\sqrt{1+\tau^2}}-1\right\}.
\end{aligned}
\label{eq:cp_bias_competing}
\end{equation}
Because $b_f<1$, reciprocal scale estimation creates positive finite-sample bias, whereas the factor $(1+\tau^2)^{-1/2}$ from measurement contamination pushes the estimate downward relative to $C_{p,X}$. A small apparent bias at one combination of $n$ and $\tau$ can therefore reflect cancellation, not reliable measurement or adequate sample information.

Pearn and Liao treat the same handoff as a capability-testing problem, developing modified lower confidence bounds and critical values for $C_{pk}$ under random gauge error and illustrating them with 70 observations \citep{pearn_liao_2005}. Unadjusted contaminated data can weaken the power of a test that capability exceeds a required level, but the correction is conditional on a supplied gauge-error quantity; it neither estimates that quantity from a finite GR\&R study nor propagates uncertainty in the gauge estimate.

This difference between a supplied parameter and an estimated measurement system recurs throughout the literature. A conditional formula answers what the decision would be \emph{if} the stated error ratio were correct, whereas a GR\&R experiment supplies an estimate with its own design, variance components, degrees of freedom, and assumptions. Celano's parametric bootstrap returns intervals for repeatability, reproducibility, and derived gage-capability measures under average--range and crossed-ANOVA designs and shows that random versus mixed appraiser models can change interval width \citep{celano_2022}. By contrast, the asymmetric-tolerance test of Grau, the precision-to-tolerance curves of Dalalah and Bani Hani, and the multivariate index $MC_p$ of Shishebori et al. condition their decisions on a supplied gage-error description \citep{grau_2011,dalalah_ptr_2016,shishebori_2009}. These studies establish different \edge{M}{C} handoffs; they do not supply a missing universal capability formula.

The wider measurement-system literature confirms that the upstream input is itself a statistical decision object. Classical designed-experiment formulations distinguish basic gauge models from variance-component estimation \citep{montgomery_runger_1993a,montgomery_runger_1993b}; misclassification-based studies attach intervals or design sizes to the risk of assigning parts incorrectly \citep{burdick_etal_2005,zappa_deldossi_2009}; and simulation or application studies show how gauge-study adequacy and within-part variation depend on the design actually used \citep{ha_etal_2017,majeske_2012}. Recent work develops a more comprehensive inferential framework for measurement-system parameters \citep{lashkari_chenouri_2025}, while a separate confidence-interval proposal illustrates continuing methodological development \abstractcitep{deldossi_zappa_2012}. These sources strengthen the \edge{M}{I} entry condition, but do not by themselves carry finite-study uncertainty through capability, monitoring, and action.

The historical literature also shows that the handoff extends beyond one index or gage criterion. Early work connected measurement-system sampling with process capability and distinguished measurement-system from process capability \citep{mcneese_1991,persijn_1996,mittag_1997}. Persijn and Van Nuland, for example, propose the source-specific rule $\mathrm{MI}=\sigma_X/\sigma_E>3$ before interpreting $C_{pk}$ and direct an inadequate result toward measurement-method improvement \citep{persijn_1996}. Later studies treat $C_{pm}$, one-sided indices, asymmetric tolerances, modified indices, precision-to-tolerance formulations, and multivariate principal-component capability \citep{bordignon_2006,pearn_2006,grau_2011,shishebori_2009,dalalah_ptr_2016,scagliarini_2011}. Pearn and Liao, for example, adjust bounds and critical values for $C_{PU}$ or $C_{PL}$ conditional on a supplied error ratio and illustrate the rule with 60 display-panel observations \citep{pearn_2006}. These are not interchangeable replications: their targets, centering losses, tolerance geometries, dimensions, and error representations differ. Collectively they establish the breadth of \edge{M}{C} and the need to retain each source's estimand and assumptions.

\subsection{Measurement criteria, improvement action, and handoff synthesis}

Several studies connect capability analysis more directly with measurement-system qualification. Al-Refaie and Bata use replicated part-by-operator observations, an analysis-of-variance decomposition, and four measures of measurement and process capability: precision-to-tolerance ratio, signal-to-noise ratio, discrimination ratio, and $C_p$ or $C_{pk}$ \citep{al_refaie_bata_2010}. The proposed sequence is operationally explicit: estimate part, operator, interaction, and repeatability components; calculate the measurement criteria; then decide whether the gage, the process, or both require improvement. Three previously studied cases illustrate the procedure.

This work is a direct procedural bridge from \stage{M} to \stage{C}, but its evidence boundary is also clear. The examples show how estimated variance components enter several threshold rules; they do not establish that sampling uncertainty in those components has been propagated through every threshold. The fixed acceptance criteria also encode different denominators and decision ideas. As noted in comparative work on gage criteria, apparently interchangeable measurement-system metrics can be algebraically related or operationally redundant while still being presented as separate checks \citep{woodall_borror_2008}. A lifecycle synthesis must therefore record which metric was used and what decision it was intended to protect rather than count the number of reported metrics as evidence strength.

Dalalah and Bani Hani formulate the actual-versus-observed $C_p$ problem through a signal-to-noise ratio \citep{dalalah_hani_2016}. Their distributions and type-I/type-II risk curves link a measurement-system description to capability acceptance. This is useful because the decision depends on specified error risks rather than on a single universal signal-to-noise cutoff. At the same time, the result remains conditional on normality, independent components, and the paper's signal-to-noise representation. The cases are illustrations drawn from the literature rather than an independent prospective validation of the acceptance rule.

The \edge{M}{C} handoff becomes a decision chain when capability is used to allocate improvement effort. Villeta et al. combine an uncertainty-to-tolerance relationship, an Index of Contamination of Capability, and a metrological improvement loop \citep{villeta_etal_2012}. Their rules distinguish situations in which measurement uncertainty is adequate for the intended capability assessment, situations in which the measurement system should be improved, and situations in which further metrological investment may be unnecessary relative to process improvement. This is best represented as \chain{M}{C}{A}: measurement uncertainty changes the interpretation of capability, and the interpreted capability then guides an action.

The study is valuable because it makes the downstream action explicit. Its four-page format and illustrative example, however, do not establish how the rule performs across processes, tolerance regimes, or alternative uncertainty budgets. It is evidence that an operational feedback rule has been proposed, not evidence that the rule is universally optimal or prospectively validated.

Selmi et al. provide another bridge by adjusting capability indices and $\bar{X}$/$R$ chart limits for measurement-system variability under specified type-I and type-II risks \citep{selmi_etal_2017}. Their piston-ring example uses 20 subgroups of five measurements. For the capability portion, the relevant handoff is \edge{M}{C}; for the adjusted limits, it is \edge{M}{II}. The two consequences share an upstream measurement-variance representation but must remain separate claims. A single industrial illustration confirms that the calculations can be carried out; it does not independently validate the corrected decisions or show that the same risk adjustment is appropriate in other industries.

The \edge{M}{C} literature establishes model-bounded mechanisms for variance contamination, adjusted confidence bounds, risk-based critical values, and joint measurement/process criteria. The remaining lifecycle issue is whether a supplied error ratio or plug-in GR\&R estimate carries its finite-study uncertainty into the capability decision. Four boundaries follow: process-only and recorded-data indices are different estimands; point adjustment, confidence bound, and shop-floor rule are not interchangeable; the source and precision of the measurement input must be reported; and improvement-allocation examples propose a \chain{M}{C}{A} loop without validating a universal policy.

For the illustrative dimensional case in Table~\ref{tab:running_case}, a crossed gage study estimates bias, repeatability, reproducibility, and their precision. These quantities calibrate the capability rule applied to recorded Phase I measurements; they do not de-noise individual part values. The rule must label its latent-process or recorded-data target and cannot reveal the source of a future signal.

\section{Measurement uncertainty and monitoring performance}
\label{sec:measurement_monitoring}

\noindent\textit{Handoff question.} Keeping the same measurement source, how does recorded-data uncertainty change false alarms, detection delay, or run length, and what action can a chart signal support before its physical cause has been identified? This branch follows \edge{M}{II}; it reaches \edge{II}{A} only when a decision rule and the information needed for attribution are also supplied.

\subsection{Measurement error and chart performance}

A control chart acts on measured observations, not on an unobserved process state. Measurement error can therefore change the distribution of the chart statistic even when the production process itself is unchanged. The downstream consequences include false-alarm behavior, detection probability, run length, and the apparent size or direction of a shift. Chart limits or statistics may be calibrated under a measurement model, but deployment still compares future recorded values with the prespecified rule; it does not replace them by reconstructed latent values. This direct \edge{M}{II} mechanism is analytically well established, although most studies evaluate it under known or calibrated error parameters rather than under a fully estimated measurement-system model.

Under the independent-replication mechanism studied for measurement-error charts \citep{linna_woodall_2001,maravelakis_panaretos_psarakis_2004}, suppose $r$ readings are averaged at time $t$ under the zero-bias version of Equation~\eqref{eq:additive_measurement_model}. Then
\begin{equation}
\overline Y_t^{(r)}=X_t+\overline E_t^{(r)},
\quad
\mathrm{Var}\!\left(\overline Y_t^{(r)}\right)
=\sigma_X^2+\frac{\sigma_E^2}{r}.
\label{eq:replicate_variance}
\end{equation}
If the process mean shifts by $\delta\sigma_X$, the shift expressed in recorded-data standard-deviation units is
\begin{equation}
\delta_{\mathrm{eff}}(r)
=\frac{\delta}{\sqrt{1+\tau^2/r}}.
\label{eq:effective_shift}
\end{equation}
This attenuation explains both the loss of detection power and the partial recovery obtained by independent replication in the reviewed studies \citep{linna_woodall_2001,maravelakis_panaretos_psarakis_2004}. It does not apply unchanged to shared bias, drift, or correlated replicate errors.

Linna and Woodall's Shewhart analysis and the EWMA/CUSUM studies provide the representative additive-error mechanism: measurement variation reduces detection power, whereas independent replication can partly restore it \citep{linna_woodall_2001,maravelakis_panaretos_psarakis_2004,maravelakis_2012}. The result is not a universal instruction to repeat every measurement because the parameters are treated as known or calibrated and the preferred replicate count depends on cost, error structure, and shift criterion.

The additive model is not the only mechanism. Abbasi studies the two-component representation \citep{abbasi_2010}
\begin{equation}
\begin{aligned}
Y_t&=a+cX_t\exp(\eta_t)+\varepsilon_t,\\
\eta_t&\sim \mathcal{N}(0,\sigma_\eta^2),
\quad \varepsilon_t\sim \mathcal{N}(0,\sigma_\varepsilon^2),
\end{aligned}
\label{eq:two_component_error}
\end{equation}
with independent multiplicative and additive disturbances. Multiplicative error becomes consequential at higher concentration while additive error remains important near zero. The studied EWMA behavior can therefore be asymmetric, and the known model parameters leave calibration-parameter estimation outside the analysis. Table~\ref{tab:monitoring_mechanisms} consolidates the wider chart-family evidence by mechanism rather than by chart name.

\begin{table*}[htbp]
\centering
\small
\caption{Representative measurement-to-monitoring mechanisms across chart families.}
\label{tab:monitoring_mechanisms}
\renewcommand{\arraystretch}{1.08}
\setlength{\tabcolsep}{4pt}
\hyphenpenalty=10000\relax
\exhyphenpenalty=10000\relax
\begin{tabularx}{\textwidth}{@{}L{0.13\textwidth}L{0.27\textwidth}L{0.26\textwidth}Y@{}}
\toprule
\textbf{Error structure} & \textbf{Representative designs} & \textbf{Main downstream effect} & \textbf{Input status and validity} \\
\midrule
Additive\newline random error & Shewhart, EWMA, CUSUM, and adaptive designs \citep{linna_woodall_2001,maravelakis_panaretos_psarakis_2004,maravelakis_2012,munir_etal_2023,ahmadini_etal_2025} & Reduced signal-to-noise ratio; replication restores detection. & Known or calibrated inputs; rankings and counts are design-specific. \\
Two-component\newline or drift & EWMA and $\bar X$ designs \citep{abbasi_2010,scagliarini_cocchi_2011,mittag_1998} & Asymmetric false alarms or detection; setup and later degradation differ. & Known parameters; calibration uncertainty and prospective action are excluded. \\
Dependent or\newline multicomponent & Autocorrelated, multivariate, ratio, and CV designs \citep{costa_2011,shongwe_2020,linna_woodall_busby_2001,amiri_etal_2018,nguyen_etal_2021,riaz_etal_2022,tran_2016} & Dependence, accuracy, precision, and correlation alter operating characteristics. & Supplied error structure; transfer across dependence regimes is source-specific. \\
Adaptive or\newline economic & Adaptive, run-rule, sampling, and economic designs \citep{daryabari_2016,hu_2014,hu_2015_vss,hu_2015_economic,javaid_2020,maleki_2021,shan_2021,haizhen_2025,abbasi_2015} & Detection, run length, sampling burden, and cost trade off. & Supplied error and cost inputs; no ranking or prospective policy. \\
\bottomrule
\end{tabularx}
\end{table*}

In a review of 60 relevant studies, Maleki et al. found that the literature was dominated by additive-error, univariate, and conventional-process settings \citep{maleki_2017_review}. Shore reverses the usual direction by deriving permissible measurement bias and variance from required Shewhart operating characteristics \citep{shore_2004}. Together, these sources show that \edge{M}{II} is an established mechanism across chart families, but the literature generally conditions on supplied or known measurement-error parameters and stops before source attribution or prospectively validated action. Effect size and chart ranking remain specific to the error structure, onset time, chart, shift, subgroup design, and performance criterion.

\subsection{From chart performance to action}

Monitoring becomes operational when it changes the measurement method, replicate count, chart, or response to a signal. Run-length and power studies address the first three more often than they validate an action policy. Replication can mitigate independent random error \citep{linna_woodall_2001,maravelakis_panaretos_psarakis_2004,maravelakis_2012,munir_etal_2023}, but it consumes capacity and may not mitigate bias, drift, autocorrelation, or operator effects. An operational choice must therefore declare measurement cost, delay, false-intervention loss, missed-shift loss, and when requalification is preferred to process adjustment.

The reviewed studies supply several bounded examples of this transition. Abbasi optimizes subgroup size and replicate count under a stated analytical-laboratory cost function, Hu et al. incorporate measurement error and Taguchi loss into an upper-sided synthetic dispersion chart, and Koszty\'an and Katona jointly optimize warning limits, control limits, variable sample sizes, and sampling intervals under a decision-cost matrix \citep{abbasi_2015,hu_2015_economic,kosztyan_2018}. Double-sampling and variable-sample-size studies likewise trade faster detection against additional observations \citep{hu_2015_vss,maleki_2021}. These studies demonstrate that cost-sensitive \chain{M}{II}{A} design is possible. Their optimum is nevertheless conditional on one chart family, a specified error model, and supplied economic inputs; none propagates uncertainty in those inputs through a prospectively validated factory policy.

The Selmi et al. procedure is unusual in connecting measurement variability to both capability values and control limits \citep{selmi_etal_2017}. It demonstrates how a risk specification can alter what is plotted and compared. It does not, however, identify whether a future signal was caused by a process change or a measurement-system change. This distinction matters: accounting for measurement error in chart design can improve the calibration of a signal, but it does not automatically solve alarm attribution.

Production measurements alone generally contain the combined consequences of process and measurement changes. Without a reference stream, replicated measurements, redundant sensing, known distinct dynamics, calibration information, or another external source, a change in the observed series need not identify its physical origin. The reviewed \edge{M}{II} studies mostly evaluate monitoring performance conditional on an error model; they do not validate a general \edge{II}{A} rule that separates process correction from measurement-system intervention.

\subsection{Monitoring-handoff synthesis}

The \edge{M}{II} evidence establishes that measurement error belongs inside chart-performance evaluation and that independent replication can mitigate random error under specific models. Improved false-alarm or run-length performance does not establish source attribution or net operational benefit. The unresolved gap concerns validated action under estimated measurement-system inputs, costs, bias, drift, and competing process changes; it is not an absence of chart variants.

For the illustrative dimensional case, replication may reduce independent random error, but a signal still combines production and measurement consequences without attribution information. The Phase II chart also inherits limits that were random when estimated in Phase I and fixed at deployment.

\section{Phase I estimation and Phase II performance}
\label{sec:phase1_phase2}

\noindent\textit{Handoff question.} Moving the source of uncertainty from the gage to the finite reference study, which Phase I quantities are estimated, how does their realized error enter the deployed chart, and which conditional or marginal guarantee is actually reported? This branch follows \edge{I}{II} and supplies the benchmark for handling estimated upstream inputs throughout the lifecycle.

\subsection{Phase I estimates change Phase II performance}

Phase II control-chart properties are often derived under known in-control parameters, whereas practice typically estimates those parameters from a Phase I reference sample. The resulting control limits are random and shared across future observations. This changes both the marginal distribution of run length across possible reference samples and the conditional performance experienced after one particular reference sample has been selected.

For illustration, the estimated-limit literature represents a $\bar X$ chart based on Phase I information $\mathcal D_I$ by random plug-in limits of the form \citep{quesenberry_1993,albers_kallenberg_2004,jensen_etal_2006}
\begin{equation}
L(\mathcal D_I)=\widehat\mu_I-K\frac{\widehat\sigma_I}{\sqrt n},
\quad
U(\mathcal D_I)=\widehat\mu_I+K\frac{\widehat\sigma_I}{\sqrt n},
\label{eq:estimated_limits}
\end{equation}
where $L(\mathcal D_I)$ and $U(\mathcal D_I)$ are the lower and upper control limits fitted from the Phase I reference data, $K$ is the chart multiplier, and $n$ is the future subgroup size. Conditional on the realized estimates, the in-control signal probability is
\begin{equation}
p_0(\mathcal D_I)=
\Pr_{\theta_0}\!\left\{\bar X_t<L(\mathcal D_I)
\ \text{or}\ \bar X_t>U(\mathcal D_I)\mid\mathcal D_I\right\}.
\label{eq:conditional_signal_probability}
\end{equation}
For independent future subgroups, the corresponding conditional average run length is $\mathrm{CARL}_0(\mathcal D_I)=1/p_0(\mathcal D_I)$. The distinction between conditional performance for one realized Phase I sample and performance averaged across reference samples is central in the estimated-parameter literature \citep{jensen_etal_2006,psarakis_etal_2013}. Across possible Phase I samples, the average conditional average run length (AARL) is
\begin{equation}
\mathrm{AARL}_0=
\mathrm{E}_{\mathcal D_I}\!\left\{\frac{1}{p_0(\mathcal D_I)}\right\}
\neq
\frac{1}{\mathrm{E}_{\mathcal D_I}\{p_0(\mathcal D_I)\}}
\quad\text{in general}.
\label{eq:conditional_marginal_arl}
\end{equation}
This distinction is the mathematical reason that an acceptable average can coexist with a non-negligible fraction of realized charts having poor false-alarm performance \citep{quesenberry_1993,albers_kallenberg_2004}.

An average alone does not state how often a deployed chart will meet a minimum false-alarm requirement. For a minimum acceptable conditional in-control average run length $L_0$ and an allowed lower-tail fraction $\alpha$ of possible Phase I reference samples, a complementary realized-chart criterion is
\begin{equation}
\Pr_{\mathcal D_I}\!\left\{
\mathrm{CARL}_0(\mathcal D_I)\ge L_0
\right\}\ge 1-\alpha,
\label{eq:realized_chart_guarantee}
\end{equation}
Equation~\eqref{eq:realized_chart_guarantee} is a canonical review representation of the percentile/exceedance logic discussed in the estimated-parameter literature, not a universal correction formula \citep{quesenberry_1993,albers_kallenberg_2004,jensen_etal_2006,psarakis_etal_2013}. An out-of-control requirement must declare its own direction, such as an upper bound on conditional detection delay, rather than reuse the in-control inequality mechanically. Thus every claimed guarantee must name the chart, estimator, Phase I information size, process state, operating quantity, target, and probability level.

Quesenberry's simulation study makes the practical consequence visible for $\bar{X}$ and individual-$X$ charts \citep{quesenberry_1993}. Estimated limits can produce a higher probability of early false alarms even in settings where the average run length increases. For the exact designs examined, the paper recommends roughly 100 subgroups of size five for permanent $\bar{X}$ limits and at least 300 observations for an individuals chart. The broader expression involving $400/(n-1)$ is presented as a speculative generalization and is not treated here as a universal sample-size rule.

Albers and Kallenberg combine asymptotic analysis and simulation to develop corrections for estimated Shewhart limits \citep{albers_kallenberg_2004}. In their normal-model setting, plug-in limits can require very large Phase I samples to approximate the intended in-control behavior. Their correction terms allow substantially smaller reference samples while retaining out-of-control sensitivity; the paper contrasts about 40 observations under the correction with an earlier recommendation near 300 for uncorrected plug-in use. These numbers belong to the paper's specific chart, model, and accuracy criterion. They demonstrate that correction is possible, not that 40 observations is adequate for every Phase I problem.

The two studies also show why average run length alone can conceal an operational problem. Estimation creates heterogeneity across realized Phase I samples: some fitted charts will be too liberal and others too conservative. A satisfactory design must state whether its guarantee is conditional on the realized estimates, averaged over possible Phase I samples, or expressed through a percentile or exceedance probability of the conditional run-length distribution.

\subsection{What the review literature establishes}

Jensen et al. review studies that explicitly evaluate control charts with estimated parameters \citep{jensen_etal_2006}. Their synthesis establishes that the issue spans multiple chart families and that small reference samples can materially distort false-signal and detection behavior. They also emphasize the need to distinguish conditional from marginal performance and suggest updating limits as stable information accumulates. This article is used here as historical and conceptual context; the primary studies it reviews are not counted again as independent replications merely because they appear in the review.

A later review by Psarakis et al. covers more than 45 studies involving Shewhart, EWMA, CUSUM, and other charts \citep{psarakis_etal_2013}. Their survey discusses efficient and robust estimators, updated limits, combined charts, multivariate settings, and additional run-length summaries. The breadth of this literature is a strong novelty constraint: a manuscript framed simply as a new review of parameter-estimation effects on control charts would duplicate an established topic. The present review uses \edge{I}{II} as one lifecycle segment and asks how its reporting and decision logic compare with measurement, capability, action, and requalification segments.

Because both articles are reviews, they serve as coverage and framing evidence. They do not provide one pooled numerical effect, and their counts cannot be combined with the present working corpus to estimate prevalence. Their strongest contribution to the lifecycle synthesis is conceptual: a chart designed with estimated inputs must be evaluated as an estimated-input procedure, not as its known-parameter analogue.

\subsection{Estimator choice as a shared upstream decision}

The Phase I handoff is not limited to the number of reference observations. The estimator used for a common process parameter can influence more than one downstream tool. Mahmoud et al. compare range-based and standard-deviation-based estimators for quality-control applications \citep{mahmoud_etal_2010}. Under normality, their bias, mean-squared-error, and relative-efficiency comparisons generally favor standard-deviation-based approaches for both $\bar{X}$ control limits and process-capability indices when subgroup size exceeds one. For equal subgroup sizes, the multiple-sample degrees of freedom are $m(n-1)$, making the available reference information explicit.

This study supports linked \edge{I}{II} and \edge{I}{C} claims because the same upstream scale estimator enters two different downstream calculations. It does not imply that capability and monitoring are the same decision. A capability index relates process behavior to specification limits, whereas a control chart evaluates temporal evidence relative to a reference state. The shared estimator creates parallel consequences; it does not erase the distinction between the destinations.

Estimator efficiency is also not the only criterion. Phase I data may contain contamination, special causes, nonnormality, autocorrelation, or model misspecification. A statistically efficient estimator under a clean normal model may be fragile under those departures. The review therefore tracks the Phase I screening assumptions, estimator, subgroup design, and degrees of freedom whenever a later claim depends on the estimated baseline.

\subsection{Why no universal Phase I sample-size rule exists}

The reviewed evidence does not support a single minimum Phase I sample size. The required reference information depends on the estimated parameters, chart family, subgroup size, in-control false-alarm criterion, desired out-of-control performance, estimator, and whether a correction is used. It also depends on the type of guarantee: matching a marginal average run length is not the same as controlling the probability that a realized chart has an unacceptably high false-alarm rate.

Table~\ref{tab:evidence_summary} summarizes both the function and credibility/applicability boundary of the main evidence used in Sections~\ref{sec:measurement_capability}--\ref{sec:phase1_phase2}. The table is a claim map, not a quality score.

\begin{table*}[htbp]
\centering
\small
\caption{What the principal handoffs establish and what remains unresolved.}
\label{tab:evidence_summary}
\renewcommand{\arraystretch}{1.10}
\setlength{\tabcolsep}{4pt}
\hyphenpenalty=10000\relax
\exhyphenpenalty=10000\relax
\begin{tabularx}{\textwidth}{@{}L{0.10\textwidth} L{0.43\textwidth} Y@{}}
\toprule
\textbf{Handoff} & \textbf{What the literature establishes} & \textbf{What remains unresolved} \\
\midrule
\edge{M}{C} & Measurement error changes capability estimates, confidence bounds, tests, and acceptance \citep{bordignon_scagliarini_2002,pearn_liao_2005,dalalah_hani_2016}. & Supplied error ratios keep the evidence model-bounded. \\
\chain{M}{C}{A} & Measurement and capability metrics can prioritize gage versus process improvement \citep{al_refaie_bata_2010,villeta_etal_2012}. & Rules are threshold-specific and lack independent prospective validation. \\
\edge{M}{II} & Measurement error changes power and run length; replication can mitigate random error \citep{linna_woodall_2001,maravelakis_panaretos_psarakis_2004,munir_etal_2023,ahmadini_etal_2025}. & Bias, cost, dependence, and attribution receive less validation. \\
\edge{I}{II} & Estimated limits alter false alarms and run length; setting-specific corrections can help \citep{quesenberry_1993,albers_kallenberg_2004,jensen_etal_2006,psarakis_etal_2013}. & Rules remain chart-, estimator-, and criterion-specific. \\
Linked\newline \edge{I}{C}/\edge{I}{II} & Scale-estimator choice can affect capability estimates and control limits \citep{mahmoud_etal_2010}. & Shared inputs do not equate conformity with stability; validation is limited. \\
\bottomrule
\end{tabularx}
\end{table*}

\subsection{Synthesis and lifecycle implication}

The broader \edge{I}{II} literature is sufficiently established that parameter-estimation effects should not be presented as a new field-wide gap. Its role here is to provide a benchmark for evidence handling: source-specific results show why the reference estimator, chart family, subgroup design, and operating criterion must travel with any sample-size or correction recommendation. The narrower synthesis here describes retained cross-stage evidence, not the maturity of the surrounding field.

In the illustrative dimensional case, the finite Phase I subgroups produce $\widehat\mu_I$ and $\widehat\sigma_I$ with their estimator and information size. The same estimates feed capability and monitoring, but their uncertainty is judged against different operating criteria.

\section{Shared Phase I estimates in capability and monitoring}
\label{sec:capability_monitoring}

\noindent\textit{Handoff question.} When the same finite, measured Phase I data feed both capability and monitoring, which uncertainty is shared, which operating consequences remain distinct, and what additional criterion turns parallel calculations into a sequential \chain{I}{C}{II} decision? The branches now rejoin, but specification conformity and temporal stability remain different destinations.

\subsection{Capability and stability are distinct}

Capability analysis and process monitoring can use the same Phase I estimates while answering different questions. Their shared-input fork is summarized by
\begin{equation}
\begin{aligned}
(\widehat\mu_I,\widehat\sigma_I)&\longrightarrow
\{\widehat C_{pk},\ \text{conformity risk}\},\\
(\widehat\mu_I,\widehat\sigma_I)&\longrightarrow
\{L(\mathcal D_I),U(\mathcal D_I),\bar X_t,\ \text{run-length risk}\}.
\end{aligned}
\label{eq:capability_monitoring_fork}
\end{equation}
Equation~\eqref{eq:capability_monitoring_fork} is a navigation representation, not a new estimator: the upper branch requires engineering specification limits, whereas the lower branch requires an in-control temporal reference and future subgroups. Both branches are computed from recorded measurements. The parameters represented by $\widehat\mu_I$ and $\widehat\sigma_I$ must nevertheless be labeled as recorded-stream or latent-process targets; the same notation must not switch targets between branches without an explicit measurement model. With a declared target, the familiar Phase I plug-in indices instantiate Equation~\eqref{eq:capability_convention_crosswalk} by setting $\mu=\widehat\mu_I$ and $s=\widehat\sigma_I$ \citep{pearn_kotz_johnson_1992,kotz_johnson_2002,anonymous_workflows_2026}:
\begin{equation}
\begin{aligned}
\widehat C_p&=\frac{\mathrm{USL}-\mathrm{LSL}}{6\widehat\sigma_I},\\
\widehat C_{pk}&=\min\left\{
\frac{\mathrm{USL}-\widehat\mu_I}{3\widehat\sigma_I},
\frac{\widehat\mu_I-\mathrm{LSL}}{3\widehat\sigma_I}
\right\}.
\end{aligned}
\label{eq:phase1_capability_estimators}
\end{equation}
Equation~\eqref{eq:estimated_limits} compares future subgroups with an estimated reference state, whereas Equation~\eqref{eq:phase1_capability_estimators} compares that state with engineering specifications. Sharing $\widehat\mu_I$ and $\widehat\sigma_I$ creates linked uncertainty, but a high capability estimate does not prove stability and an in-control chart does not prove conformance.

The capability-restricted and acceptance-chart designs of Oprime et al. make the sequential connection explicit by allowing a capability criterion to constrain an $\bar X$ chart and by using the sampling distributions of $\widehat C_p$ and $\widehat C_{pk}$ in economic design \citep{oprime_2017,oprime_etal_2019}. This is properly coded as \chain{I}{C}{II}: Phase I estimates determine a capability statement, and the capability criterion helps determine the monitoring or acceptance design. The practical examples show how the optimization can be implemented, but they do not convert every conventional control chart into a capability decision or establish transportability across cost structures and distributions. Workflows that place capability indices inside a statistical process control (SPC) sequence provide useful implementation context while remaining adjacent unless the uncertainty transfer itself is evaluated \citep{hrncirikova_2024,anonymous_workflows_2026}.

\subsection{From finite-sample capability to decision risk}

The sampling uncertainty of the capability estimate creates an \edge{I}{C} decision problem even when measurement error is absent. Using the specification midpoint $m$ defined in Section~\ref{sec:measurement_capability}, let $h=(\mathrm{USL}-\mathrm{LSL})/2$, so that $C_{pk}=\{h-|\mu-m|\}/(3\sigma)$ under the normal-process definition. Combining classical capability-inference results and the risk-calibrated approval study \citep{pearn_kotz_johnson_1992,anonymous_risk_2026} with the delta-method formulation \abstractcitep{oehlert_1992} and specialized $C_{pk}$ asymptotic theory \abstractcitep{chen_pearn_1997} gives the following local representation only when $\mu\neq m$, the nearer specification side is locally unique, and the joint estimator of $\mu$ and $\sigma$ satisfies the required regular limit:
\begin{equation}
\sqrt{n}\left(\widehat C_{pk}-C_{pk}\right)
\ \dot\sim\ \mathcal{N}(0,\sigma_C^2),
\label{eq:cpk_local_asymptotic}
\end{equation}
where $\sigma_C^2$ is the source-specific asymptotic variance in that regular off-center case. At exact centering, the absolute-value/minimum functional is non-differentiable in $\mu$; the ordinary normal delta-method statement is therefore not asserted, and a centered-case distribution or exact source-specific method is required. For an approval threshold $C_0$, the regular off-center representation gives the following local approximation, where $\Phi$ is the standard normal cumulative distribution function:
\begin{equation}
\Pr\!\left(\widehat C_{pk}\ge C_0\mid C_{pk}\right)
\approx
\Phi\!\left\{\frac{\sqrt n(C_{pk}-C_0)}{\sigma_C}\right\}.
\label{eq:capability_approval_probability}
\end{equation}
Under the same regular off-center approximation, the point rule $\widehat C_{pk}\ge C_0$ has approximately one-half approval probability at the boundary. This statement is not transferred to the centered non-differentiable case. A risk-calibrated guard band instead uses
\begin{equation}
\widehat C_{pk}\ge C_0+k\,\widehat{\mathrm{SE}}(\widehat C_{pk}),
\label{eq:capability_guard_band}
\end{equation}
where $\widehat{\mathrm{SE}}$ denotes the estimated standard error and $k$ is calibrated to a tolerated false-approval probability or an explicit loss ratio \citep{anonymous_risk_2026}. This decision rule clarifies the inferential-to-decision handoff. It remains a rule applied to an estimate obtained from recorded data. It does not automatically include measurement-system uncertainty unless that uncertainty is represented in $\widehat{\mathrm{SE}}(\widehat C_{pk})$ or in the decision model, and it does not authorize replacing the measured part values by putative error-free values.

Recent capability-based decision-support work extends this logic from a single threshold to structured predictive decisions \citep{anonymous_hybrid_2026}. Its value for the review is two-sided: an uncertainty-aware structure can make risk components interpretable, but the reported grouped evaluation did not show universal predictive superiority over simpler logistic baselines. The appropriate synthesis is therefore that structure and predictive discrimination are separate validation targets; complexity is not evidence of better transfer.

\subsection{Uncertainty enters twice when measured data define both stages}

When the Phase I observations are measured with error, the same recorded variance may enter both Equations~\eqref{eq:estimated_limits} and \eqref{eq:phase1_capability_estimators}. The capability value is then affected by the distinction between $\sigma_X$ and $\sqrt{\sigma_X^2+\sigma_E^2}$, while the monitoring limits inherit uncertainty from the estimated combined scale. A procedure may be internally consistent for recorded-data monitoring yet still fail to estimate process-only capability. Conversely, subtracting an estimated measurement variance can define a model-based population estimate of $\sigma_X^2$ for a process-only capability target, but it neither recovers $X_i$ from each $Y_i$ nor automatically provides correctly calibrated monitoring limits. The deployed chart must still be designed for and applied to the future recorded stream.

Selmi et al. demonstrate a procedure that adjusts both capability indices and $\bar X$/$R$ limits under a specified measurement-error and risk formulation \citep{selmi_etal_2017}. Ben Amara et al. likewise derive Shewhart $\bar X$ and $R$ limits calibrated under an explicit measurement-error model while examining capability, nonconformance, and loss quantities \abstractcitep{ben_amara_etal_2016}. These are direct \stage{M--C--II} precedents: the common input creates coordinated calculations, but capability and monitoring retain different decision consequences. Their examples establish computational feasibility; they do not show that a future signal can be attributed uniquely to the production process or that the joint adjustment is optimal when measurement-system inputs are themselves uncertain.

\subsection{Evidence boundary for the sequential chain}

The reviewed evidence supports the existence of capability-constrained monitoring designs, not a universal rule for combining capability and control charts. A transferable study must report the specification target, Phase I estimator and information size, measurement-error treatment, chart family, false-acceptance and false-alarm criteria, cost assumptions, and whether validation concerns calculations, simulation, retrospective data, or prospective decisions. Without those elements, placing a capability index beside a control chart is an adjacent workflow rather than evidence of uncertainty propagation.

For the illustrative dimensional case, Equation~\eqref{eq:capability_monitoring_fork} sends estimates derived from the same recorded Phase I values to two destinations: LSL/USL define the capability branch, whereas the temporal reference and future recorded subgroups define the monitoring branch. A process-only capability target may use a separately identified variance decomposition, but the Phase II chart continues to observe $Y_t$. Either branch can yield a calibrated statistical decision without identifying what physically changed. Closing the case requires an action rule, source-discriminating information, and a post-action check.

\section{From statistical signals to attributed actions}
\label{sec:action_requalification}

\noindent\textit{Handoff question.} Given a capability result or Phase II signal, what action is taken, what loss or operating criterion selects it, and what independent information can verify whether the intervention addressed the production process, the measurement system, or both? This final handoff follows \edge{II}{A} and \edge{A}{R}; unlike the preceding branches, it cannot be completed by a statistical signal alone.

\subsection{Action requires an explicit decision rule}

The destination \stage{A} requires a defined intervention: release, stop, investigate, repeat measurement, adjust the process, recalibrate the instrument, or collect new reference information. Most reviewed studies end with bias, coverage, power, or run length. Those quantities can inform action, but they do not by themselves establish which intervention minimizes operational loss or whether the intervention corrected the true source of degradation.

Villeta et al. provide one explicit \chain{M}{C}{A} rule through their Index of Contamination of Capability (ICC) \citep{villeta_etal_2012},
\begin{equation}
\mathrm{ICC}=100\frac{\widehat C_{p,\mathrm{obs}}}
{\widehat C_{p,\mathrm{real}}}
=100\frac{\widehat C_{pk,\mathrm{obs}}}
{\widehat C_{pk,\mathrm{real}}}.
\label{eq:icc}
\end{equation}
The equality of the two displayed ratios is source-specific: it requires the same process spread in the paired observed/real indices and a centering/bias construction under which the $C_{pk}$ numerator is affected compatibly with the $C_p$ comparison. It should not be transferred to a biased or differently centered process merely because both indices are available. Under the source model and operational construction, ICC values below 58 prioritize improvement of the measurement system, values from 58 to 91 treat it as adequate for the intended capability evaluation, and values above 91 indicate potentially excessive measurement resources. These cutoffs are source-specific resource-allocation rules derived from the paper's uncertainty-to-tolerance assumptions and illustrative setting; they do not convert production readings into error-free part values and are not universal loss-optimal thresholds.

\subsection{Requalification requires discriminating evidence}

Requalification asks whether the measurement system, uncertainty budget, or reference state remains fit for use after time, intervention, or process improvement. Song and Vorburger use a check-standard history and dynamic limits to test whether selected components of a measurement-uncertainty budget remain credible \citep{song_vorburger_2007}. Risk- and reliability-based calibration-interval studies provide adjacent examples of turning drift history, financial exposure, or instrument classifications into a recalibration schedule \abstractcitep{pashnina_2018,janatyan_etal_2025}. These support the need for an explicit \edge{M}{R} decision object, but their setting-specific cost, aging, and classification models are not universal requalification policies. A reference artifact provides information that ordinary production observations do not contain; passing one chart does not demonstrate that all biases are absent.

The information requirement is substantive. In the additive representation $Y_t=X_t+E_t$, a change in $Y_t$ alone can be explained by a change in $X_t$, a change in $E_t$, or both. Replicates can estimate short-term random error under independence, but they need not identify common drift. Check standards, calibration artifacts, redundant measurement methods, or deliberately separated dynamics can make attribution more credible. Without such information, a production alarm can justify investigation but not a unique choice between process adjustment and measurement-system requalification.

\begin{figure*}[htbp]
\centering
\resizebox{\textwidth}{!}{%
\begin{tikzpicture}[
  font=\sffamily\small,
  head/.style={rounded corners=3pt,minimum width=4.15cm,minimum height=0.84cm,
    text width=3.85cm,align=center,inner sep=3pt,font=\sffamily\small\bfseries,text=white},
  body/.style={rounded corners=3pt,minimum width=4.15cm,minimum height=2.55cm,
    text width=3.75cm,align=center,inner sep=5pt},
  verify/.style={rounded corners=3pt,draw=stagecoral!75,fill=stagecoral!7,
    minimum width=9.3cm,minimum height=0.72cm,text width=8.9cm,align=center},
  arr/.style={-{Stealth[length=2.1mm]},draw=stageink,line width=0.9pt},
  feedback/.style={-{Stealth[length=2.1mm]},draw=stagecoral,line width=0.9pt,dashed}
]
\node[head,fill=stageblue] at (0,3.15) {Competing\\sources};
\node[head,fill=stagecoral] at (4.75,3.15) {Recorded\\signal};
\node[head,fill=stageteal] at (9.50,3.15) {Discriminating\\evidence};
\node[head,fill=stagegold] at (14.25,3.15) {Attributed\\action};

\node[body,draw=stageblue!70,fill=stageblue!8] (causes) at (0,1.25)
  {Process change: $\Delta X_t\neq0$\\[2pt]
   Gage change: $\Delta E_t\neq0$\\[2pt]
   Both sources may change};
\node[body,draw=stagecoral!85,fill=stagecoral!10] (alarm) at (4.75,1.25)
  {$\Delta Y_t=\Delta X_t+\Delta E_t$\\[4pt]
   Signal $\Rightarrow$ investigate\\[2pt]
   Cause remains unresolved};
\node[body,draw=stageteal!80,fill=stageteal!8] (info) at (9.50,1.25)
  {Reference artifact\\[2pt]
   Replication or redundancy\\[2pt]
   Model, residual, or intervention record};
\node[body,draw=stagegold!85,fill=stagegold!10] (act) at (14.25,1.25)
  {Contain or adjust\\[2pt]
   Requalify the system\\[2pt]
   Verify the post-action state};

\draw[arr] (causes.east) -- (alarm.west);
\draw[arr] (alarm.east) -- (info.west);
\draw[arr] (info.east) -- (act.west);
\node[verify] (post) at (7.12,-0.64)
  {Post-action evidence renews the measurement or process baseline.};
\draw[feedback] (act.south) |- (post.east);
\draw[feedback] (post.west) -| (causes.south);
\end{tikzpicture}
}
\caption{From alarm to attributed action. A recorded signal can have competing process and measurement sources; attribution therefore requires reference, redundancy, model, or intervention information, followed by post-action verification. No single diagnostic architecture is assumed to be universally sufficient.}
\label{fig:attribution_evidence}
\end{figure*}
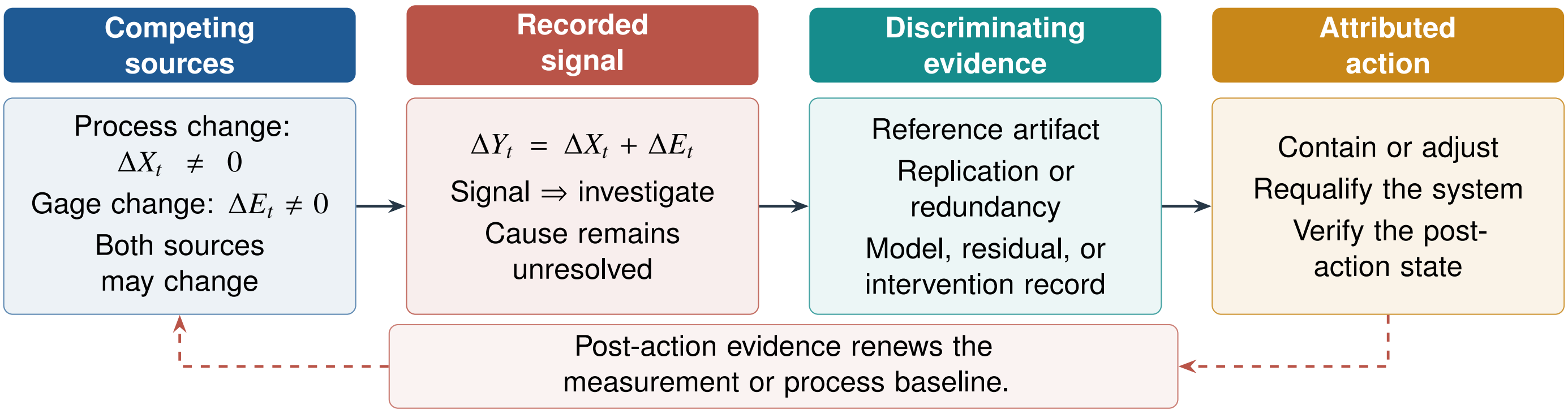

Figure~\ref{fig:attribution_evidence} makes the identification boundary explicit. Its left side shows why one production series does not uniquely identify the physical source of an alarm; its right side summarizes the additional information structures used in check-standard, reconstruction, residual, and sensor-fusion studies \citep{song_vorburger_2007,doymaz_2001,cho_2012,salahshoor_2008,park_2020}.

The adjacent fault-detection literature demonstrates what such discriminating information can look like. Model-based residuals, Kalman filtering, and centralized or decentralized data fusion have been used to separate sensor failures or drift from process upsets \citep{doymaz_2001,cho_2012,salahshoor_2008}, while broader process-diagnostics reviews organize the available fault-detection architectures \citep{park_2020}. Complementary publisher-abstract reviews distinguish quantitative models, qualitative causal/search representations, process-history methods, model-based diagnosis, and data-driven identifiability \abstractcitep{venkatasubramanian_etal_2003a,venkatasubramanian_etal_2003b,venkatasubramanian_etal_2003c,isermann_2005,qin_2012}. Cho and Jiang, for example, model a ramp sensor drift as an augmented Kalman-filter state and use a nonzero innovation mean for detection and drift-rate estimation in a simulated pressurizer \citep{cho_2012}. The result depends on an adequate dynamic model, operating-point linearization, sampling rate, and observation window; the paper does not supply prospective field validation or the confidence analysis for choosing that window. These sources are relevant to \edge{II}{A} because they make attribution an explicit inference problem. They are not direct replications of univariate capability or control-chart studies, and their richer state, redundancy, or dynamic-model assumptions must be preserved.

Doymaz et al. add a complementary information structure: paired $T^2$ and squared-prediction-error statistics, neighbor-sensor correlation, and reconstruction from redundant measurements are used to distinguish failed sensors from process upsets in an industrial melter benchmark \citep{doymaz_2001}. Salahshoor et al. instead augment adaptive extended-Kalman-filter states with sensor bias and process-fault terms and compare centralized with decentralized fusion in a simulated nonlinear reactor \citep{salahshoor_2008}. Both studies show why attribution requires information beyond a single production series. They also mark the transfer boundary: linear correlation, model adequacy, redundancy, and simulation fidelity replace the simpler assumptions of univariate chart studies, and neither constitutes prospective validation of a generic requalification rule.

Risk-based conformity assessment provides one bridge from statistical output to action by attaching losses or producer/consumer risks to decisions \citep{desimoni_2011,forbes_2006,separovic_lourenco_2020}. Risk-based chart design embeds related decision consequences in chart selection and adjustment \citep{kosztyan_2016,kosztyan_2017,kosztyan_2018}. Economic chart design has long made sampling, alarm, and shift consequences explicit \abstractcitep{duncan_1956,lorenzen_vance_1986}; a centenary perspective nevertheless cautions that formal economic designs have limited documented incorporation into routine chart practice \abstractcitep{montgomery_2024}. For the four acceptance/rejection outcomes used by Koszty\'an et al., let $K_L,K_U$ be the acceptance-limit corrections, $D_{K_L,K_U}(Y)$ the recorded-data decision, $C(X)$ the actual conformity state, and $\pi_{ij}$ the payoff for decision $i$ and state $j$, with $i,j\in\{0,1\}$. The expected payoff $\Pi$ is
\begin{equation}
\begin{aligned}
\Pi(K_L,K_U)&=\sum_{i=0}^{1}\sum_{j=0}^{1}\pi_{ij}
\Pr\{D_{K_L,K_U}(Y)=i,\ C(X)=j\},\\
(K_L^\star,K_U^\star)&=\arg\max_{K_L,K_U}\Pi(K_L,K_U).
\end{aligned}
\label{eq:conformity_profit}
\end{equation}
Equation~\eqref{eq:conformity_profit} displays the industrially consistent distinction directly \citep{kosztyan_2017}: the unobserved conformity state concerns $X$, while the implemented acceptance rule acts on $Y$ and uses measurement information to choose its boundaries. The correction changes the rule, not the recorded observation. The equation also shows why an uncertainty interval alone does not determine an action: the same statistical information can produce different corrections under different payoff ratios. The source assumes known real-value and measurement-error distributions and a fixed one-dimensional payoff structure; it does not establish one end-to-end policy for every manufacturing lifecycle.

\subsection{Closing the loop and its validation boundary}

A complete feedback claim would connect a signal to a prespecified action, document the information used to attribute the cause, and evaluate the post-action result. The retained literature contains proposed loops and illustrative decisions, but little independent prospective evidence comparing alternative interventions. The defensible conclusion is therefore that feedback architectures exist while validated \edge{II}{A} and \edge{A}{R} policies remain comparatively sparse in the evidence examined here. This is an evidence gap in transfer and validation, not proof that industry never performs these actions.

For the illustrative dimensional case, a signal therefore initiates investigation rather than identifying an intervention. Replicates, a check standard, redundant sensing, calibration evidence, or distinguishable dynamics must separate a process change from a gage change; a declared loss or operating rule then selects the action. Requalification closes the case only when post-action evidence renews the affected gage or process baseline. The following synthesis evaluates every handoff against this same sequence.

\section{Evidence strength, transferability, and gaps}
\label{sec:evidence_strength}

\subsection{Evidence strength and transfer conditions}

Within the retained literature, measurement-to-capability and measurement-to-monitoring handoffs are more extensively represented by model-based studies. Direct Phase I-to-capability evidence is less extensively represented and remains estimator-, chart-, and criterion-specific, so it is not treated as independently replicated. Prospective evidence is thinner when a signal must be attributed to a physical source, converted into a cost-sensitive intervention, and followed by requalification. The reviewed action and requalification studies provide mechanisms and illustrations, but little independent prospective operational validation. Reviews broaden coverage and standards describe intended practice, but neither substitutes for such evaluation.

An equation transfers only with its input, estimand, and assumptions. The attenuation in Equation~\eqref{eq:observed_capability} requires an additive independent model and a clear distinction between latent-process and recorded-data capability; it does not imply that the latent realization is observed. The replication gain in Equation~\eqref{eq:effective_shift} requires independent random replicate errors. The run-length distinction in Equation~\eqref{eq:conditional_marginal_arl} requires recognition that Phase I estimates remain fixed for the future chart after being randomly selected and that the deployed chart still receives measured observations. None of these results alone supplies a loss function, source-attribution rule, or requalification schedule.

The same caution applies to evidence types. Analytical derivation can establish a mechanism; simulation can quantify it over a chosen grid; an industrial illustration can establish feasibility; and prospective multi-setting evaluation can test transfer. The retained corpus is strongest in the first two functions and thinner in the latter two. Review articles broaden coverage but are not additional primary replications, and standards describe intended practice without estimating the operating effect of a particular lifecycle handoff.

\subsection{Evidence layers and review contribution}

The evidence does not support a universal lifecycle equation. It supports an audit structure: identify the upstream uncertain object and its information status; specify its propagation; name the downstream decision and performance criterion; and preserve the source's transfer boundary. Equations~\eqref{eq:additive_measurement_model}--\eqref{eq:capability_guard_band} make recurring mechanisms explicit without asserting that all lifecycle nodes follow one probability model.

Each handoff is read through three layers. The \emph{physical and measurement layer} explains how recorded observations arise; the \emph{inferential layer} names the latent or recorded target and propagation of uncertainty; and the \emph{decision and validation layer} states the rule, action, loss or operating criterion, and supporting evidence. Strength in one layer does not imply strength in another: an exact run-length calculation need not identify a signal's cause, and an industrial workflow need not provide calibrated false-approval risk.

The major cross-stage gap is limited validation of multistage decisions when measurement-system quantities are estimated, Phase I information is finite, causes compete, and costs are unequal. The illustrative dimensional case can compute measurement components, capability quantities, fitted limits, and operating characteristics, but it requires additional information and post-action evidence to attribute a signal or validate an intervention. This evidence map, rather than a universal estimator, is the review's contribution.

\section{Research agenda for lifecycle evidence}
\label{sec:agenda}

The synthesis identifies five research needs that are narrower than inventing another isolated estimator. First, studies should report a common upstream uncertainty record: the recorded input, targeted estimand, information size, model, covariance or interval, and whether each measurement-system quantity was known, calibrated, or estimated. Second, work that connects measurement and Phase I estimation should evaluate their joint effect on rules applied to recorded data rather than correcting one component while treating the other as fixed or implying observation-level error removal. Third, monitoring designs should report conditional and marginal operating characteristics and, when one realized Phase I estimate will remain fixed in deployment, a percentile or exceedance guarantee for the realized chart.

Fourth, action studies need source-discriminating information from replicates, standards, redundant sensors, or identifiable dynamics. Fifth, prospective evaluation should compare prespecified actions under unequal costs and document the post-action requalification outcome. Across these priorities, digital implementations should test versioned uncertainty records as explicit inputs to capability, monitoring, release, and requalification procedures; this is an engineering research opportunity, not an established standard.

Synthesizing the outcome domains used in capability inference, monitoring, decision-risk, and requalification studies \citep{pearn_liao_2005,jensen_etal_2006,forbes_2006,song_vorburger_2007}, an end-to-end evaluation should report the review-level vector
\begin{equation}
\mathcal V=
\begin{gathered}
(\text{bias/coverage},\ \text{false approval},\\
\text{false alarm/delay},\ \text{action loss},\\
\text{requalification outcome}).
\end{gathered}
\label{eq:evaluation_vector}
\end{equation}
The components of $\mathcal V$ need not be combined into one score. Their purpose is to prevent success at an upstream statistical metric from being silently relabeled as deployment or physical-mechanism validation. A useful new study may target only part of the vector, provided that its evidence boundary is explicit.

\section{Conclusion}
\label{sec:conclusion}

Measurement-system information is an upstream decision input, not a detachable prelude to manufacturing quality analysis. Its status, whether known, calibrated, or estimated, must remain visible when recorded observations feed capability inference, Phase I reference estimation, Phase II monitoring, action, and requalification. Measurement uncertainty should calibrate downstream limits, uncertainty statements, replication, guard bands, and losses; it should not be interpreted as observation-level de-noising or as permission to replace the production record by an unobserved latent value.

The retained evidence repeatedly documents model-based consequences for capability and monitoring and the effect of estimated Phase I inputs on deployed chart performance. It is less extensive where a signal must be attributed to a process or measurement source, converted into a cost-sensitive action, and verified through requalification. The handoff representation makes those boundaries explicit without proposing a universal estimator or claiming database-complete coverage. Operationally, a downstream analysis should identify the version and validity domain of the measurement evidence it consumed; a monitoring signal should initiate investigation rather than automatic causal attribution; and post-action evidence should renew the affected measurement record, process reference, and downstream consumers. The literature examined here does not establish a universally validated end-to-end policy. The resulting opportunity for measurement science is to design and prospectively test calibration artifacts, replication, redundancy, dynamic models, and versioned uncertainty records that preserve uncertainty status and enable defensible downstream attribution and requalification.

\section*{DECLARATIONS}
\begin{description}[itemsep=0pt,parsep=0pt,topsep=2pt,partopsep=0pt]
\item[Data availability:] No new experimental or manufacturing data were generated or analyzed in this review.
\item[Code availability:] Not applicable. No analysis or simulation code was used or generated in this review.
\item[Ethics approval:] Not applicable.
\item[Consent to participate:] Not applicable.
\end{description}

\section*{Generative AI disclosure}
ChatGPT (OpenAI, GPT-5) was used solely for language polishing; all scientific content was verified and approved by the authors.

\bibliographystyle{elsarticle-num}
\bibliography{references}

\end{document}